\PassOptionsToPackage{table}{xcolor}
\documentclass[a4paper,12pt]{article}
\usepackage{jheppub} % for details on the use of the package, please
\usepackage[T1]{fontenc} % if needed

\usepackage[percent]{overpic}
\usepackage{slashed}
\usepackage{wrapfig}
\usepackage{tabu}
\usepackage{mathrsfs,amsmath,amssymb,amsthm,amsfonts,tikz,graphicx,accents,hyperref, color}
\usepackage{dsfont,epiolmec, latexsym, stmaryrd, comment}
\usepackage{slashed,ccaption}
\usepackage{mathrsfs, calligra}
\usepackage{leftidx}
\usepackage{import}
\usepackage{multirow}
\usepackage{amsfonts}
\usepackage{pifont}
\usepackage{tabularx}
\usepackage[utf8]{inputenc}
\usetikzlibrary{intersections,calc}
\usepackage{tikz-3dplot}
\usepackage{ifthen}
\usetikzlibrary{arrows}
\usepackage{amsmath}
\usepackage{xcolor}
\usepackage{geometry}

\usepackage{array}
\DeclareFontFamily{OT1}{rsfs}{}

\DeclareFontShape{OT1}{rsfs}{m}{n}{ <-7> rsfs5 <7-10> rsfs7 <10->rsfs10}{} 

\DeclareMathAlphabet{\mycal}{OT1}{rsfs}{m}{n}

\newcommand{\bms}{{$\mathfrak{bms}_3$}}
\newcommand{\hbms}{$\widehat{\mathfrak{bms}}_3$}

\newcommand{\Max}{$\mathfrak{Max}_{3}$}

\newcommand{\eps}{\varepsilon}

\newcommand{\be}{\begin{equation}}
\newcommand{\ee}{\end{equation}}

\tdplotsetmaincoords{48}{112}
\makeatletter \@addtoreset{equation}{section}

\newcommand\hnote[1]{\textcolor{magenta}{\bf [Hamid:\,#1]}}

\preprint{IPM/P-2019/037}

\title{{\boldmath \centerline{\emph{On Stabilization of Maxwell-BMS Algebra}}}}

\author[a]{P. Concha}
\author[b]{and H. R. Safari}
\affiliation{$^a$ Departamento de Matemática y Física aplicadas, Universidad Católica de la Santísima Concepción, Alonso de Ribera 2850, Concepción, Chile}
\affiliation{$^b$ School of Physics, Institute for Research in Fundamental
Sciences (IPM),\\ P.O.Box 19395-5531, Tehran, Iran}

\emailAdd{patrick.concha@ucsc.cl}
\emailAdd{hrsafari@ipm.ir}
\abstract{

In this work we present different infinite dimensional algebras which appear as deformations of the asymptotic symmetry of the three-dimensional Chern-Simons gravity for the Maxwell algebra. We study rigidity and stability of the infinite dimensional enhancement of the Maxwell algebra. In particular, we show that three copies of the Witt algebra and the $\mathfrak{bms}_{3}\oplus\mathfrak{witt}$ algebra are obtained by deforming its ideal part. New family of infinite dimensional algebras are obtained by considering deformations of the other commutators which we have denoted as $M(a,b;c,d)$ and $\bar{M}(\bar{\alpha},\bar{\beta};\bar{\nu})$. Interestingly, for the specific values $a=c=d=0, b=-\frac{1}{2}$ the obtained algebra $M(0,-\frac{1}{2};0,0)$ corresponds to the twisted Schr\"{o}dinger-Virasoro algebra. The central extensions of our results are also explored. The physical implications and relevance of the deformed algebras introduced here are discussed along the work.}

\begin{document}
\maketitle

\section{Introduction and motivations}

Symmetry is the cornerstone of the modern theoretical physics. Among different symmetries, the symmetries of spacetimes have attracted more attentions. One particular symmetry is the Poincar\'e algebra  which is isometry of Minkowski spacetime and field theories on flat space enjoy Poincar\'e invariance. Depending on the theory and its field content, field theories typically exhibit invariance under bigger symmetry algebras which can be seen as extensions and deformations of the Poincar\'e algebra.

A well-known extension and deformation of the Poincar\'e algebra is given by the Maxwell algebra which is characterized by the presence of an Abelian anti-symmetric two tensor generators $\mathcal{M}_{\mu\nu}$ such that the generators of translations obey $[\mathcal{P}_{\mu},\mathcal{P}_{\nu}]=\mathcal{M}_{\mu\nu}$. This algebra was first introduced in \cite{schrader1972maxwell, beckers1983minimal} where it describes a particle in an external constant electromagnetic field background, see also \cite{Bacry:1970ye, Soroka:2004fj}. This algebra can be obtained from the study of Chevalley-Eilenberg cohomology of Poincar\'e algebra \cite{Bonanos:2008kr,Bonanos:2008ez}. In the context of the gravity by gauging the $4d$ Maxwell algebra an extension of General Relativity (GR) is obtained which includes a generalized cosmological term \cite{deAzcarraga:2010sw}. Subsequently, in the context of three dimensional gravitational theories, a Chern-Simons (CS) gravity theory invariant under $2+1$ Maxwell algebra was studied in \cite{Salgado:2014jka,Hoseinzadeh:2014bla,Concha:2018zeb}. In three spacetime dimensions, an isomorphic (dual) version of the Maxwell algebra, denoted as Hietarinta-Maxwell algebra, has been useful in the study of spontaneous symmetry breaking \cite{Bansal:2018qyz}. Remarkably, both topological and minimal massive gravity theories \cite{Deser:1981wh, Bergshoeff:2014pca} can be seen as particular cases of a more general minimal massive gravity arising from a spontaneous breaking of a local symmetry in a Hietarinta-Maxwell CS theory \cite{Chernyavsky:2020fqs}.

Another class of extensions of the Poincar\'e algebra in $3d$ and $4d$ to infinite dimensional algebras appear in the asymptotic symmetry algebra analysis where respectively $\mathfrak{bms}_3$ and $\mathfrak{bms}_4$ algebras are obtained \cite{Bondi:1962px, Sachs:1962zza, Sachs:1962wk, Barnich:2006av, Barnich:2009se, Barnich:2011ct, Barnich:2011mi, Barnich:2012aw}. In these algebras $3d$ and $4d$  Poincar\'e algebra appear as the maximal global part of the algebra. There has been a renewed interest in these asymptotic symmetries as they could be used to provide an alternative derivation for the Weinberg's soft theorems as well as the memory effect \cite{Strominger:2014pwa,Strominger:2017zoo,Pate:2019mfs}.

One may then ask if $3d$ or $4d$ Maxwell algebras also admit a similar infinite dimensional enhancement. The answer to this question is affirmative. In particular, there exist the \Max\ algebra which is an infinite dimensional enhancement of  the $2+1$ Maxwell algebra \cite{Caroca:2017onr}. Interestingly, the \Max\ algebra can be obtained as an extension and deformation of the $\mathfrak{bms}_3$ \cite{Parsa:2018kys}.
Moreover, it has been shown that a centrally extended version of \Max\ algebra also arises in the asymptotic symmetry analysis of certain $3d$ Maxwell CS gravity theory \cite{Concha:2018zeb}. In particular, as in GR, the geometries described by the field equations of the Maxwell CS theory are Riemannian and locally flat. However, the so-called gravitational Maxwell field couples to the geometry leads to particular effects different to those of GR. Indeed, it modifies not only the asymptotic sector but also the vacuum energy and momentum of the stationary configuration. Furthermore the vacuum energy, unlike in GR where it is always non zero \cite{Barnich:2012aw,Miskovic:2016mvs}, can be vanished for particular values of the coupling constant of the Maxwell CS term.  %In particular, this theory admits a Minkowski vacuum solution with non-vanishing two form field associated with other fields. This latter is then related to the fact that \Max\ algebra has $3d$ Poincar\'e as its subalgebra \cite{Barnich:2012aw, Miskovic:2016mvs}.
 
As a next natural question, one may ask if there are other infinite dimensional algebras arising as asymptotic symmetries. The answer is ``Yes''. The asymptotic symmetry algebras strongly depend on the choice of the boundary conditions. For instance, if one chooses the boundary condition on near horizon of a black holes instead of asymptotic infinity region, one obtains a completely different symmetry algebra as it has recently been shown in \cite{Grumiller:2019fmp} that the symmetry algebra in near horizon of $3d$ and $4d$ black holes yields $W(0,b)$ or $W(a,a;a,a)$ algebras, see \cite{Parsa:2018kys, Safari:2019zmc} for the definition of the latter. Given that there are practically infinitely many possibilities for boundary conditions and infinitely many surfaces (like null infinity or horizon) to impose boundary conditions over, there seems to be infinitely many such infinite dimensional algebras. Hence, a natural question is whether one can find/classify other such algebras?

To answer this question one may rely on ``algebraic'' techniques rather than the asymptotic symmetry analysis. One approach may be provided through deformation theory of Lie algebras where given any Lie algebra one can in principle deform it to obtain another algebra. The procedure of deformation can be continued until we reach a rigid (stable) algebra, which cannot be further deformed. For finite dimensional Lie algebras there is the Hochschild-Serre factorization (HSF) theorem \cite{MR0054581} which sets the stage: The end process of deformation of an algebra is a semi-simple Lie algebra with the same dimension. For example, $d$ dimensional Poincar\'e algebra is not stable and it could be deformed to $\mathfrak{so}(d-2,2)$ or $\mathfrak{so}(d-1,1)$ algebras. The question of stability/deformation of infinite dimensional algebras has not yet been tackled in full generality. There are some case-by-case analysis, e.g. see  \cite{Parsa:2018kys, Safari:2019zmc}. For example, the $\mathfrak{bms}_{3}$ can be deformed to two copies of Virasoro algebras, which is asymptotic symmetry of $AdS_{3}$ spacetime, or $W(0,b)$ which is symmetry algebra of near horizon of $3d$ black holes \cite{Grumiller:2019fmp}. Also by starting with $\mathfrak{bms}_{4}$, one shows that it can be deformed into $W(a,a;a,a)$ which symmetry algebra of near horizon of $4d$ black holes.
 
Motivated by the diverse applications of the Maxwell algebra and by the recent results obtained through deformation of asymptotic symmetries,here we explore deformations of the infinite dimensional extension of the 3d Maxwell algebra given by the \Max\ algebra. We find different new infinite dimensional algebras which may potentially appear as asymptotic/near horizon symmetry of certain physical theories with specific boundary conditions. For instance, we show that \Max\ algebra can be deformed to three copies of Virasoro algebras which has been obtained as asymptotic symmetry of another Chern-Simons gravity theory \cite{Concha:2018jjj}. Also we find that the \Max\ algebra can be deformed to a four parameter family algebra that we call $M(a,b;c,d)$ where for specific value of parameters it is the asymptotic symmetry of Schr\"{o}dinger spacetimes \cite{Compere:2009qm}. The central extension of the new algebra obtained through deformation of \Max\ algebra is also considered.

\paragraph{Organization of the paper.} In section \ref{sec:2}, we review the Maxwell algebra, its infinite dimensional enhancement in $2+1$ dimension and its deformations. In section \ref{sec:3}, we analyse various deformations of the infinite dimensional enhancement of the Maxwell algebra by studying the most general deformation of the \Max\ algebra. In section \ref{sec:4}, we study the central extensions of the obtained algebras through deformations of the infinite dimensional Maxwell algebra. Finally we summarize our results and discuss their physical interpretations.%\hnote{This part should be revised.} %In appendix \ref{deformation-theory} we review some basic concepts of deformation theory of Lie algebras.

\paragraph{Notation.} Following \cite{Oblak:2016eij} we use ``\emph{fraktur} fonts'' for algebras e.g. \bms, $\mathfrak{bms}_{4}$, \Max\ and their centrally extended versions will be denoted by a hat \hbms, $\widehat{\mathfrak{bms}}_4$ and $\widehat{\mathfrak{Max}}_3$. We also denote two family algebras $M(a,b;c,d)$ and $\bar{M}(\bar{\alpha},\bar{\beta};\bar{\nu})$ which in our conventions $\mathfrak{Max}_3=M(0,-1;0,-1)=\bar{M}(0,0;0)$. On the other hand, “$M(a,b;c,d)$ family” of algebras (of $M(a,b;c,d)$ family, in short), shall denote set of algebras for different values of the $a,b,c$ and $d$ parameters and similarly for $\bar{M}(\bar{\alpha},\bar{\beta};\bar{\nu})$ family.

 %%%%%%%%%%%%%%%%%%%%%%%%%%%%%%%%%%%%%%%%%%%%%%%%%%%%%%%%%%%%%%%%%%%%%%%%%%%%%%%%%%%%%%%%%%%%%%%%%%%%%%%%%%%%%%%

\section {Maxwell algebra and its infinite dimensional enhancement}\label{sec:2}

In this section we briefly review the  Maxwell algebra, its deformations and its infinite dimensional enhancement in $2+1$ spacetime dimensions. The discussion about how such infinite dimensional algebra can be obtained as extension and deformation of \bms\ algebra is also presented.

\subsection{The Maxwell algebra}
The Maxwell algebra in $d$ dimension can be obtained as an extension and deformation of the Poincar\'{e} algebra. In fact one can extend the Poincar\'{e} algebra by adding Lorentz-covariant tensors which are abelian as follows
\begin{equation}
    [\mathcal{J}_{\mu\nu},\mathcal{M}_{\alpha\beta}]=-(\eta_{\alpha[\mu}\mathcal{M}_{\nu]\beta}-\eta_{\beta[\mu}\mathcal{M}_{\nu]\alpha}),
\end{equation}
where $\mathcal{J}_{\mu\nu}$ are generators of the Lorentz algebra $\mathfrak{so}(d-1,1)$. Furthermore, one can deform the commutator of translations so that it is no more zero but proportional to the new generators $\mathcal{M}$ to obtain Maxwell algebra as
\begin{equation}
\begin{split}
&[\mathcal{J}_{\mu\nu},\mathcal{J}_{\alpha\beta}]=-(\eta_{\alpha[\mu}\mathcal{J}_{\nu]\beta}-\eta_{\beta[\mu}\mathcal{J}_{\nu]\alpha}),\\
&[\mathcal{J}_{\mu\nu},\mathcal{P}_{\alpha}]=-(\eta_{\mu\alpha}\mathcal{P}_{\nu}-\eta_{\nu\alpha}\mathcal{P}_{\mu}),\\
   & [\mathcal{J}_{\mu\nu},\mathcal{M}_{\alpha\beta}]=-(\eta_{\alpha[\mu}\mathcal{M}_{\nu]\beta}-\eta_{\beta[\mu}\mathcal{M}_{\nu]\alpha}),\\
   &    [\mathcal{P}_{\mu},\mathcal{P}_{\nu}]=\varepsilon \mathcal{M}_{\mu\nu},
\end{split}
\end{equation}
where $\varepsilon$ is the deformation parameter. As we have mentioned this algebra describes a relativistic particle which is coupled to a constant electromagnetic field \cite{schrader1972maxwell,beckers1983minimal} and has subsequently been studied in the gravity context by diverse authors in \cite{Bonanos:2009wy,Izaurieta:2009hz,Lukierski:2010dy,Durka:2011nf,deAzcarraga:2012qj,Concha:2013uhq,deAzcarraga:2014jpa,Concha:2014tca,Concha:2014vka,Concha:2014zsa,Cebecioglu:2015jta,Concha:2016zdb,Caroca:2017izc,Ravera:2018vra,Aviles:2018jzw,Concha:2018jxx,Kibaroglu:2018mcn,Concha:2019icz,Barducci:2019fjc,Chernyavsky:2019hyp,Concha:2019lhn}. In three spacetime dimensions, the Poincar\'{e} algebra has six generators, three generators for rotation and boost and three generators for translation. In the $3d$ Maxwell algebra, the Lorentz-covariant tensor adds three independent generators. Thus the Maxwell algebra in three spacetime dimensions has $9$ generators which can be written in an appropriate basis ($\mathfrak{sl}(2,\mathbb{R})$ basis) as
\begin{equation} 
\begin{split}\label{maxwell}
 & i[\mathcal{J}_m,\mathcal{J}_n]=(m-n)\mathcal{J}_{m+n}, \\
 &i[\mathcal{J}_m,\mathcal{P}_n]=(m-n)\mathcal{P}_{m+n},\\
 &i[\mathcal{J}_m,\mathcal{M}_n]=(m-n)\mathcal{M}_{m+n},\\
 &i[\mathcal{P}_m,\mathcal{P}_n]=(m-n)\mathcal{M}_{m+n},
\end{split}
\end{equation}
where $m,n=\pm1,0$. One then shows that the $3d$ Maxwell algebra can be enlarged to a new algebra with countable basis where $m,n\in \mathbb{Z}$ \cite{Caroca:2017onr}. In this work we shall denote the infinite dimensional version of the Maxwell algebra by $\mathfrak{Max}_{3}$. Interestingly, as was shown in \cite{Concha:2018zeb}, the latter can be obtained as the  asymptotic symmetry of a $3d$ Chern-Simons gravity based on the Maxwell algebra.

\subsection{Infinite dimensional \texorpdfstring{$3d$}{3d} Maxwell algebra through \texorpdfstring{$\mathfrak{bms}_{3}$}{bms3} algebra}

An infinite dimensional enhancement of $3d$ Maxwell algebra $\mathfrak{Max}_{3}$ can be obtained as an extension and deformation of the \bms\ algebra. Let us review  properties of the \bms\ algebra.

\paragraph{The \bms\ algebra.}

The \bms\ algebra is the centerless asymptotic symmetry of three-dimensional flat spacetime \cite{Ashtekar:1996cd, Barnich:2006av}:
 \begin{equation} 
\begin{split}
 & i[\mathcal{J}_m,\mathcal{J}_n]=(m-n)\mathcal{J}_{m+n}, \\
 &i[\mathcal{J}_m,\mathcal{P}_n]=(m-n)\mathcal{P}_{m+n},\\
 &i[\mathcal{P}_m,\mathcal{P}_n]=0,
\end{split}\label{bms3}
\end{equation}
where $m,n\in \mathbb{Z}$. The $\mathfrak{bms}_{3}$ algebra is an infinite dimensional algebra which contains two sets of generators given by ${\cal J}_n$ and ${\cal P}_n$. ${\cal J}$ generates a Witt subalgebra of \bms\ which is the algebra of smooth vector fields on a circle. On the other hand ${\cal P}_n$  generates an adjoint representation of the Witt algebra and form the ideal part of the \bms\ algebra. From \eqref{bms3} one can see that $\mathfrak{bms}_{3}$ has a semi-direct sum structure:
\begin{equation} \label{bms=witt+ideal}
\mathfrak{bms}_{3}= \mathfrak{witt}\inplus_{ad}\mathfrak{witt}_{ab},
\end{equation}
where the subscript $ab$ is to emphasize the abelian nature of ${\cal P}$ while $ad$ denotes the adjoint action.  
The maximal finite subalgebra of $\mathfrak{bms}_{3}$ is the three dimensional Poincar\'{e} algebra $\mathfrak{iso}(2,1)$, associated with restricting $m, n=\pm 1, 0$ in relation (\ref{bms3}).
In particular the generators ${\cal J}$ and ${\cal P}$ are called superrotations and supertranslations, respectively.

A central extension of the \bms\ algebra, denoted as $\widehat{\mathfrak{bms}}_{3}$, appears by asymptotic symmetry analysis of three dimensional flat space:
 \begin{equation}\label{BMS-centrally-extended} 
\begin{split}
 & i[\mathcal{J}_m,\mathcal{J}_n]=(m-n)\mathcal{J}_{m+n}+\frac{c_{JJ}}{12}m^{3}\delta_{m+n,0}, \\
 &i[\mathcal{J}_m,\mathcal{P}_n]=(m-n)\mathcal{P}_{m+n}+\frac{c_{JP}}{12}m^{3}\delta_{m+n,0},\\
 &i[\mathcal{P}_m,\mathcal{P}_n]=0,
\end{split}
\end{equation}
in which $c_{JJ}$ and $c_{JP}$ are the central charges and are related to the coupling constants of the so-called exotic Lagrangian and the Einstein-Hilbert Lagrangian as follows \cite{Barnich:2006av, Barnich:2014cwa}
\begin{equation}
\begin{split}\label{CentralTerms1}
&c_{JJ}=12k\alpha_{0}, \\
&c_{JP}=12k\alpha_{1},
\end{split}
\end{equation}
Note that the central part can also contain a term proportional to $m$. However, this part can be absorbed into a shift of   generators by a central term.

\paragraph {Extension of \bms\ algebra. }
We are interested in a particular extension of the \bms\ algebra, denoted by ${\widetilde{\mathfrak{bms}}}_{3}$, in which the additional generators have the same conformal weight as the \bms\ generators, $h=2$. The non vanishing commutators of $\widetilde{\mathfrak{bms}}_{3}$ are given by
\begin{equation} 
\begin{split}
 & i[\mathcal{J}_m,\mathcal{J}_n]=(m-n)\mathcal{J}_{m+n}, \\
 &i[\mathcal{J}_m,\mathcal{P}_n]=(m-n)\mathcal{P}_{m+n},\\
 &i[\mathcal{J}_m,\mathcal{M}_n]=(m-n)\mathcal{M}_{m+n},
\end{split}\label{bms3-extended}
\end{equation}
in which $m,\ n\ \in \mathbb{Z}$, and is defined over the field  $\mathbb{R}$. One can see that the algebra \eqref{bms3-extended} has a Witt subalgebra. In particular, the structure of $\widetilde{\mathfrak{bms}}_{3}$ is the semi direct sum of the Witt algebra with an abelian ideal part. The latter is the direct sum of generators $\mathcal{P}$ and $\mathcal{M}$. Then, we have
\begin{equation} 
\widetilde{\mathfrak{bms}}_{3}= \mathfrak{witt}\inplus (\mathfrak{P}\oplus\mathfrak{M})_{ab},
\end{equation}
where the $\mathfrak{P}$ and $\mathfrak{M}$ abelian ideals are spanned by $\mathcal{P}$ and $\mathcal{M}$ generators, respectively. 
One can show that $\widetilde{\mathfrak{bms}}_{3}$ admits only three independent central terms as 
\begin{equation} 
\begin{split}
 & i[\mathcal{J}_m,\mathcal{J}_n]=(m-n)\mathcal{J}_{m+n}+\frac{c_{JJ}}{12}m^{3}\delta_{m+n,0}, \\
 &i[\mathcal{J}_m,\mathcal{P}_n]=(m-n)\mathcal{P}_{m+n}+\frac{c_{JP}}{12}m^{3}\delta_{m+n,0},\\
 &i[\mathcal{J}_m,\mathcal{M}_n]=(m-n)\mathcal{M}_{m+n}+\frac{c_{JM}}{12}m^{3}\delta_{m+n,0},
\end{split}\label{bms3-central-extended}
\end{equation}

One can deform the  algebra in \eqref{bms3-extended} to obtain a new non isomorphic algebra with non vanishing commutators similarly to \eqref{maxwell}.
Thus one can view the \Max\ algebra \eqref{maxwell} as an extension and deformation of the \bms\ algebra. 
 The \Max\ algebra as the centrally extended $\widetilde{\mathfrak{bms}}_{3}$ algebra \eqref{bms3-extended} admits only three independent central terms as
\begin{equation} 
\begin{split}\label{C-maxwell}
 & i[\mathcal{J}_m,\mathcal{J}_n]=(m-n)\mathcal{J}_{m+n}+\frac{c_{JJ}}{12}m^{3}\delta_{m+n,0}, \\
 &i[\mathcal{J}_m,\mathcal{P}_n]=(m-n)\mathcal{P}_{m+n}+\frac{c_{JP}}{12}m^{3}\delta_{m+n,0},\\
 &i[\mathcal{J}_m,\mathcal{M}_n]=(m-n)\mathcal{M}_{m+n}+\frac{c_{JM}}{12}m^{3}\delta_{m+n,0},\\
 &i[\mathcal{P}_m,\mathcal{P}_n]=(m-n)\mathcal{M}_{m+n}+\frac{c_{JM}}{12}m^{3}\delta_{m+n,0}.
\end{split}
\end{equation}
We denote the central extension of \Max\ by $\widehat{\mathfrak{Max}}_{3}$ with the commutators as \eqref{C-maxwell}.

Such infinite-dimensional symmetry algebra in presence of three central terms can also be obtained through the semi-group expansion method \cite{Caroca:2017onr}. This algebra describes the asymptotic symmetry of a three-dimensional Chern-Simons gravity theory invariant under the Maxwell algebra \cite{Concha:2018zeb}. Interestingly, the central charges $c_{JJ}$, $c_{JP}$ and $c_{JM}$ can be related to three terms of the Chern-Simons Maxwell gravity action as follows \cite{Concha:2018zeb}:
\begin{equation}
\begin{split}\label{CentralTerms2}
c_{JJ}=12k\alpha_{0}, \qquad c_{JP}=12k\alpha_{1}, \qquad c_{JM}=12k\alpha_{2},
\end{split}
\end{equation}
where $\alpha_{0}$, $\alpha_{1}$ and $\alpha_{2}$ are the coupling constants of the exotic Lagrangian, the Einstein-Hilbert term and the so-called Gravitational Maxwell Lagrangian, respectively.

%\subsection{Review on deformation of the Maxwell algebra}

 %%%%%%%%%%%%%%%%%%%%%%%%%%%%%%%%%%%%%%%%%%%%%%%%%%%%%%%%%%%%%%%%%%%%%%%%%%%%%%%%%%%%%%%%%%%%%%%%%%%%%%%%%%%%%%

 %%%%%%%%%%%%%%%%%%%%%%%%%%%%%%%%%%%%%%%%%%%%%%%%%%%%%%%%%%%%%%%%%%%%%%%%%%%%%%%%%%%%%%%%%%%%%%%%%%%%%%%%%%%%%%%%%%%%%%%%%%%%%%%

%-------------------------------------------------------------------------------------------------------------------------------------------------------------------
\section{Deformation of \texorpdfstring{$\mathfrak{Max}_{3}$}{max3} algebra}\label{sec:3}
In this section we study deformation of the $\mathfrak{Max}_{3}$ algebra defined through \eqref{maxwell}. At the finite-dimensional level, deformations of the Maxwell algebra has been considered in \cite{Gomis:2009dm} leading to non-isomorphic algebras. In particular, the Maxwell algebra can be deformed in arbitrary dimensions to the $\mathfrak{so}(d-1,2)\oplus\mathfrak{so}(d-1,1)$ and , $\mathfrak{so}(d,1)\oplus\mathfrak{so}(d-1,1)$ algebra. The former is the direct sum of $AdS_d$ and $d$-dimensional Lorentz algebra and was studied in \cite{Soroka:2006aj, Salgado:2014qqa}. In specific dimension $d=2+1$ a the Maxwell algebra can also be deformed to the $\mathfrak{iso}(2,1)\oplus\mathfrak{so}(2,1)$ algebra.
As discussed in \cite{Parsa:2018kys} the infinite dimensional Lie algebras are not subject to Hochschild-Serre factorization theorem. Therefore, unlike the finite dimensional case, their deformations should be studied case-by-case. Here we can, not only,  deform the ideal part, but also the other commutators. We explore possible deformations of the $\mathfrak{Max}_{3}$ algebra by deforming  all commutators simultaneously. Then, we explore which of the previous infinitesimal deformations are also a formal deformation. As it is discussed in \cite{Parsa:2018kys} there are different ways to show that an infinitesimal deformation is formal. As was pointed out in \cite{Parsa:2018kys}, ``the quick test'' is the approach we apply here in which one shows that the infinitesimal solution can satisfy the Jacobi identities for all order of the deformation parameter.   Specific cases where only some commutators are deformed are also discussed. We also provide an algebraic cohomology analysis. 

Further details about deformation and stability can be found in \cite{Parsa:2018kys} where an exhaustive description of deformation of Lie algebras has been presented.

%------------------------------------------------------------------------------------------------------------------------------------------------------------------------------------------------
%\subsection{Deformation of  \bms\ algebra}
%------------------------------------------------------------------------------------------------------------------------------------------------------------------------------------------------

%%%%%%%%%%%%%%%%%%%%%%%%%%%%%%%%%%%%%%%%%%%%%%%%%%%%%%%%%%%%%%%%%%%%%%%%%%%%%%%%%%%%%%%%%%%%%%%%%%
\subsection{The most general deformation of \texorpdfstring{$\mathfrak{Max}_{3}$}{Max3} algebra}\label{sec:3.4}

To our purpose we consider all deformations of the commutators of the \Max\ algebra. The most general deformation of the \Max\ algebra is given by:
\begin{equation} 
\begin{split}\label{most-general-deform}
&i[\mathcal{J}_{m},\mathcal{J}_{n}]=(m-n)\mathcal{J}_{m+n}+(m-n)F(m,n)\mathcal{P}_{m+n}+(m-n)G(m,n)\mathcal{M}_{m+n},\\
&i[\mathcal{J}_{m},\mathcal{P}_{n}]=(m-n)\mathcal{P}_{m+n}+K(m,n)\mathcal{P}_{m+n}+I(m,n)\mathcal{M}_{m+n}+O(m,n)\mathcal{J}_{m+n},\\
&i[\mathcal{J}_{m},\mathcal{M}_{n}]=(m-n)\mathcal{M}_{m+n}+\tilde{K}(m,n)\mathcal{M}_{m+n}+\tilde{I}(m,n)\mathcal{P}_{m+n}+\tilde{O}(m,n)\mathcal{J}_{m+n},\\
 &i[\mathcal{P}_m,\mathcal{P}_n]=(m-n)\mathcal{M}_{m+n}+(m-n)f_{1}(m,n)\mathcal{P}_{m+n}+(m-n)h_{1}(m,n)\mathcal{J}_{m+n}, \\
 &i[\mathcal{P}_m,\mathcal{M}_n]=f_{2}(m,n)\mathcal{P}_{m+n}+g_{2}(m,n)\mathcal{M}_{m+n}+h_{2}(m,n)\mathcal{J}_{m+n},\\
 &i[\mathcal{M}_m,\mathcal{M}_n]=(m-n)f_{3}(m,n)\mathcal{P}_{m+n}+(m-n)g_{3}(m,n)\mathcal{M}_{m+n}+(m-n)h_{3}(m,n)\mathcal{J}_{m+n},
\end{split}
\end{equation}

where the arbitrary functions can be fixed from the Jacobi identities leading to diverse deformations. It is important to emphasize that throughout this work the indices of the generators $\mathcal{J}$, $\mathcal{P}$ and $\mathcal{M}$ which appear in the right-hand-side are fixed to be $m+n$. On the other hand, the functions have a polynomial expansion in term of their arguments. Furthermore, we shall not write the deformation term as $(m-n)g_{1}(m,n)\mathcal{M}_{m+n}$ which just rescales the term $(m-n)\mathcal{M}_{m+n}$ by a constant parameter as $\alpha(m-n)\mathcal{M}_{m+n}$. Of course this can be absorbed into a redefinition of generators. In what follows we study each Jacobi identity and its respective implications.

\subsection*{Infinitesimal deformation}
In this part we study the deformation which is called "{\it infinitesimal}" deformation in which we consider the constraints obtained from the Jacobi identities in first order of the functions. 
Let us consider first the Jacobi identity $[\mathcal{J},[\mathcal{J},\mathcal{J}]]+\text{cyclic permutations}=0$ which implies in the first order in functions:
\begin{multline} \label{F,G-eq}
(n-l)(m-n-l)[G(m,l+n)+G(n,l)]+ (l-m)(n-l-m)[G(n,l+m)+G(l,m)]+\\
 (m-n)(l-m-n)[G(l,m+n)+G(m,n)]=0.
\end{multline}
Analogously, the same relation will be obtained for $F(m,n)$. In particular, there is no other constraint for $G$. Then, we have:
\begin{equation}\label{JJ-Z}
        G(m,n)=Z(m)+Z(n)-Z(m+n),
\end{equation}
provides a solution to \eqref{F,G-eq}, for any arbitrary function $Z$ and can be seen as the most general solution. Nevertheless, it is possible to show that the deformations of the form \eqref{JJ-Z} are trivial deformations since they can be reabsorbed by redefining the generators\footnote{One can show that this result is true when different deformations (functions) are turned on simultaneously.} as: 
\begin{equation} 
\begin{split}\label{ZY-redefinition}
  \mathcal{J}_{m}:=\tilde{\mathcal{J}}_{m}+Z(m)\tilde{\mathcal{M}}_{m}, \qquad \mathcal{P}_{m}:=\tilde{\mathcal{P}}_{m}, \qquad \mathcal{M}_{m}:=\tilde{\mathcal{M}}_{m},
\end{split}
\end{equation}
where $\tilde{\mathcal{J}}_{m}$, $\tilde{\mathcal{P}}_{m}$ and $\tilde{\mathcal{M}}_{m}$ satisfy the commutation relations of the \Max\ algebra \eqref{maxwell}.

On the other hand, one finds from the Jacobi identity $[\mathcal{J},[\mathcal{J},\mathcal{P}]]+\text{cyclic permutations}=0$ the following relation at the first order for $K$:
  \begin{multline}\label{firsteq-K}
(n-l) K(m,l+n)+(m-n-l)K(n,l)+(l-m) K(n,l+m) +\\
+(l+m-n)K(m,l)+(n-m)K(m+n,l)=0,
\end{multline}
which can be solved, as was discussed in \cite{Parsa:2018kys}, by
\begin{equation} \label{KIO-solution}
 K(m,n)=\alpha+\beta m+\gamma m(m-n)+\cdots.
\end{equation}
The same relation and solution is found for $O$. One can see that the Jacobi identity $[\mathcal{J},[\mathcal{J},\mathcal{M}]]+\text{cyclic permutations}=0$ also leads to the same relation and solution as \eqref{firsteq-K} and \eqref{KIO-solution} for $\tilde{K},\tilde{I}$ and $\tilde{O}$.

Furthermore, the Jacobi identity $[\mathcal{J},[\mathcal{J},\mathcal{P}]]+\text{cyclic permutations}=0$ also leads to a relation for functions $F$ and $I$ as
 \begin{multline}\label{I-F}
   (n-l) I(m,l+n)+(m-n-l)I(n,l)+(l-m)I(n,l+m) +\\
+(l+m-n)I(m,l)+(n-m)I(m+n,l)+(m-n)(l-m-n)F(m,n)=0.
\end{multline}
which is solved by $I(m,n)=\bar{\alpha}+\bar{\beta}m-\bar{\nu} n+(\bar{\gamma} mn^{2}+\frac{1}{2}(\bar{\lambda}-\bar{\gamma})nm^{2}+\frac{1}{2}(-\bar{\lambda}-\bar{\gamma})m^{3})+\cdots$ and $F(m,n)=\bar{\nu}+\bar{\lambda} mn+\cdots$.

One can see that three independent relations appear by considering the Jacobi identity $[\mathcal{J},[\mathcal{P},\mathcal{M}]]$  $+\text{cyclic permutations}=0$ in the first order in functions.  In particular, we have the following relation for $h_{2}$:
\begin{equation}\label{f,g,h-2(1)}
(n-l)h_{2}(m,l+n)+(m-l)h_{2}(m+l,n)+(l-m-n)h_{2}(m,n)=0.
\end{equation}
By setting $m=n=l$ we obtain $mh_{2}(m,m)=0$. Then we have that $h_{2}(m,m)=0$ for $m\neq 0$. This means that we can write $h_{2}(m,n)=(m-n)\bar{h}_{2}(m,n)$ where $\bar{h}_{2}(m,n)$ is a symmetric function. By inserting the latter into \eqref{f,g,h-2(1)} one gets
\begin{equation}\label{f,g,h-2(2)}
(n-l)(m-l-n)\bar{h}_{2}(m,l+n)+(l-m)(n-m-l)\bar{h}_{2}(m+l,n)+(l-m-n)(m-n)\bar{h}_{2}(m,n)=0,  
\end{equation}
which is solved for $h_{2}(m,n)=\alpha(m-n)$ where $\alpha$ is arbitrary constant.
For the functions $f_{2}$ and $\tilde{O}$ one obtains
\begin{equation}\label{f2-tildeO}
 (n-l)f_{2}(m,l+n)+(m-l)f_{2}(m+l,n)+(l-m-n)f_{2}(m,n)-(m-n-l)\tilde{O}(l,n)=0.
  \end{equation}
Then, by replacing $m=n+l$ one finds the same relation as \eqref{f,g,h-2(1)} leading to $f_{2}(m,n)=\beta(m-n)$ and $\tilde{O}(m,n)=0$.
Furthermore the Jacobi identity $[\mathcal{J},[\mathcal{P},\mathcal{M}]]+\text{cyclic permutations}=0$ gives rise to a relation for $g_{2}, O$ and $\tilde{I}$ as follows
\begin{multline}\label{g2-O-tildeI}
    (n-l)g_{2}(m,l+n)+(m-l)g_{2}(m+l,n)+(l-m-n)g_{2}(m,n)+\\ 
    +(n+l-m)\tilde{I}(l,n)+(n-l-m)O(l,m)=0.
\end{multline}

By studying the Jacobi identity $[\mathcal{J},[\mathcal{P},\mathcal{P}]]+\text{cyclic permutations}=0$ it is possible to see that such identity puts not only the following constraints on the functions
\begin{equation} 
\begin{split}\label{K-tildeK,O-tildeI}
     &(n+l-m)K(l,n)+(n-l-m)K(l,m)+(m-n)\tilde{K}(l,m+n)=0,\\
& (n+l-m)O(l,n)+(n-l-m)O(l,m)+(m-n)\tilde{I}(l,m+n)=0.
\end{split}
\end{equation}
but also leads to a new relation for $f_{1}, O$ and $\tilde{I}$ as
\begin{multline}\label{f1-O-tildeI}
    (n-l)(m-n-l)f_{1}(m,l+n)+ (l-m)(n-l-m)f_{1}(n,l+m)+(m-n)(l-m-n)f_{1}(m,n)+ \\ 
    +(n+l-m)O(l,n)+(n-l-m)O(l,m)+(m-n)\tilde{I}(l,m+n)=0.
\end{multline}

One may note that the relation \eqref{f1-O-tildeI} is linear in $f_{1}, O$ and $\tilde{I}$. Furthermore, the coefficients appearing along the $O$ and $\tilde{I}$ terms are first order in $m,n,l$ while the coefficients of the $f_{1}$ terms are second order in $m,n,l$. We expect that these functions are polynomials of positive powers in their arguments, so one concludes that if $O$ and $\tilde{I}$ are monomials of degree $p$ we have that $f_{1}$ should be a monomial of degree $p+1$. Since the solutions of $O$ and $\tilde{I}$ are similar to the ones of \eqref{KIO-solution}, we have that \eqref{f1-O-tildeI} is satisfied considering $f_{1}(m,n)=constant$, $O(m,n)=\alpha+\beta m+\gamma m(m-n)$ and $\tilde{I}=2\alpha+2\beta m +\tilde{\gamma}m(m-n)$. On the other hand, one finds that \eqref{g2-O-tildeI} is linear in all functions so they should appear as monomial with the same degree. Then one can insert the solutions $O(m,n)=\alpha+\beta m+\gamma m(m-n)$ and $\tilde{I}(m,n)=2\alpha+2\beta m +\tilde{\gamma}m(m-n)$ into \eqref{g2-O-tildeI} and finds that there is no solution for $g_{2}(m,n)$ for none of them. Thus we have to set $g_{2}(m,n)=0$, which implies that $O(m,n)=\tilde{I}(m,n)=0$.

In the case of $h_{1}$, one can find a relation for such function from the Jacobi identity $[\mathcal{P},[\mathcal{P},\mathcal{P}]]+\text{cyclic permutations}=0$ which implies at first order in function the following relation:
\begin{equation}\label{f1h1-2}
(n-l)(m-n-l)h_{1}(n,l)+ (l-m)(n-l-m)h_{1}(l,m)+(m-n)(l-m-n)h_{1}(m,n) =0,
\end{equation}
which is solved for $h_{1}(m,n)=constant$. 

Following the same procedure, it is possible to show from the Jacobi identity $[\mathcal{P},[\mathcal{P},\mathcal{M}]]+\text{cyclic permutations}=0$ that the functions $g_{3}$, $f_{3}$ and $h_{3}$ have to satisfy:
 \begin{multline}\label{f2-g3-h1}
(m-n-l)f_{2}(n,l)-(n-m-l)f_{2}(m,l)+(m-n)(l-m-n)g_{3}(l,m+n)+
\\ (m-n)(l-m-n)h_{1}(m,n)=0,
\end{multline}
 \begin{equation} 
\label{h2-f3}
 (m-n-l)h_{2}(n,l)-(n-m-l)h_{2}(m,l)+(m-n)(l-m-n)f_{3}(l,m+n)=0,\\
\end{equation} 
and 
\begin{equation}\label{h3}
    (m-n)(l-m-n)h_{3}(l,m+n)=0,
\end{equation}
which imply $h_{3}(m,n)=0,g_{3}(m,n)=\text{constant}$ and $f_{3}(m,n)=\text{constant}$.

Finally, one can show that the Jacobi identity $[\mathcal{J},[\mathcal{M},\mathcal{M}]]+ \text{cyclic}$ just leads to the same results as before for $f_{3}, g_{3}$ and $h_{3}$, while the Jacobi identity $[\mathcal{M},[\mathcal{M},\mathcal{M}]]+ \text{cyclic}$ does not lead to any new constraint in first order of functions.

%For other coefficients or higher order terms in $O$ and $\tilde{I}$ there is no solution for $f_{1}$.

%The other Jacobi identity to consider is $[\mathcal{J},[\mathcal{J},\mathcal{P}]]+\text{cyclic permutations}=0$ which leads to 
%\begin{multline}\label{I-F}
 %  (n-l) I(m,l+n)+(m-n-l)I(n,l)+(l-m)I(n,l+m) +\\
%(l+m-n)I(m,l)+(n-m)I(m+n,l)+(m-n)(l-m-n)F(m,n)=0.
%\end{multline}
%By inserting the solution \eqref{JJ-Z} into \eqref{I-F} one finds that $I(m,n)=\bar{\alpha}+\bar{\beta}-\bar{\nu} n+(\bar{\gamma} mn^{2}+\frac{1}{2}(\bar{\lambda}-\bar{\gamma})nm^{2}+\frac{1}{2}(-\bar{\lambda}-\bar{\gamma})m^{3})+...$ where we have assumed $F(m,n)=\bar{\nu}+\bar{\lambda} mn+...$. Otherwise, if one sets $F(m,n)=0$ one would obtain the same solution as \eqref{KIO-solution} for $I(m,n)$. 

\subsection*{Formal deformation}

Until here we have obtained non trivial solutions for different functions which led to simultaneous infinitesimal deformations. 
In fact, we can turn on infinitesimally the functions $f_{1}$, $f_{2}$, $f_{3}$, $g_{2}$, $g_{3}$, $h_{1}$, $h_{2}$, $K$, $\tilde{K}$, $I$, $G$ and $F$ at the same time. However all of these infinitesimal deformations can not be extended to a "{\it formal}" deformations. To obtain a formal deformation, the functions should satisfy the Jacobi identities for all orders in functions. Here without entering the details, we will review the possible formal deformations.

As summary, one can see from the Jacobi identities that the non-trivial formal deformations of the \Max\ algebra can be classified in four different algebras. As we shall see, two of the deformed algebras can be written as the direct sum of known structures. The others deformed algebras are new infinite dimensional algebras. In particular, a new family algebra reproduces, for particular values, interesting results already known in the literature. In what follows we discuss the diverse deformations obtained induced by one of several functions simultaneously. One can show that there is no additional formal deformations when we consider other possible infinitesimal deformations induced by the present functions.

\subsubsection{The \texorpdfstring{$\bar{M}(\bar{\alpha},\bar{\beta};\bar{\nu})$}{M(bar-alpha,bar-beta)} algebra}

One of the new formal deformations obtained is induced by the functions $F(m,n)=\bar{\nu}$ and $I(m,n)=\bar{\alpha}+\bar{\beta}m-\bar{\nu} n$ coming from \eqref{F-I-JJJ} such that the new algebra satisfies the following non vanishing commutation relations:
\begin{equation} 
\begin{split}\label{new-family-IF}
 & i[\mathcal{J}_m,\mathcal{J}_n]=(m-n)\mathcal{J}_{m+n}+\bar{\nu}(m-n)\mathcal{P}_{m+n}, \\
 &i[\mathcal{J}_m,\mathcal{P}_n]=(m-n)\mathcal{P}_{m+n}+(\bar{\alpha}+\bar{\beta}m-\bar{\nu} n)\mathcal{M}_{m+n},\\
 &i[\mathcal{J}_m,\mathcal{M}_n]=(m-n)\mathcal{M}_{m+n},\\
 &i[\mathcal{P}_m,\mathcal{P}_n]=(m-n)\mathcal{M}_{m+n}.
\end{split}
\end{equation}
We call this new family algebra as $\bar{M}(\bar{\alpha},\bar{\beta};\bar{\nu})$. 
 One can check that the functions $F(m,n)$ and $I(m,n)$ are fixed by the Jacobi identity $[\mathcal{J},[\mathcal{J},\mathcal{J}]]+\text{cyclic permutations}=0$ which implies the non linear relation in deformation parameter as
\begin{multline}\label{F-I-JJJ}
    (n-l)F(n,l)I(m,n+l)+(l-m)F(l,m)I(n,l+m)+(m-n)F(m,n)I(l,m+n)=0,
\end{multline}
whose solution is given by $F(m,n)=\bar{\nu}$ and $I(m,n)=\bar{\alpha}+\bar{\beta}m-\bar{\nu}n$.

To our knowledge, this is a novel structure whose global part has not been explored yet. It would be interesting to study the implication of such symmetry and analyze diverse values for $\bar{\alpha}$, $\bar{\beta}$ and $\bar{\nu}$.

It is interesting to note that $\bar{\nu}=0$ reproduces a deformed algebra induced by $I=\bar{\alpha}-\bar{\beta}m$. The particular case $\bar{M}(\bar{\alpha},\bar{\beta};0)$ can be recovered by deforming only the commutator $[\mathcal{J}_m,\mathcal{P}_n]$ which implies $I=\bar{\alpha}+\bar{\beta}m+\bar{\gamma}m(m-n)+\cdots$ from the Jacobi identity $[\mathcal{J},[\mathcal{J},\mathcal{P}]]+\text{cyclic permutations}=0$ as we have previously discussed. A specific redefinition of the generators can be considered as
\begin{equation} 
\begin{split}\label{redefine-I}
  \mathcal{J}_m\equiv \tilde{\mathcal{J}}_m, \qquad \mathcal{P}_m\equiv \tilde{\mathcal{P}}_m+F(m)\tilde{\mathcal{M}}_m, \qquad \mathcal{M}_m\equiv \tilde{\mathcal{M}}_m.
\end{split}
\end{equation}
This redefinition does not change the ideal part and yields to the following relation:
\begin{equation}
    (m-n)(F(n)-F(m+n))\tilde{\mathcal{M}}_{m+n}=I(m,n)\tilde{\mathcal{M}}_{m+n}.
\end{equation}
One can then check that the solutions given by $I(m,n)=\bar{\gamma} m(m-n)+...$ can be absorbed by the above redefinition when $F(m)=a_{0} +a_{1} m+a_{2}m^{2}+\cdots$. In this way, the only non trivial formal deformation induced by $I(m,n)$ is 
\begin{equation}\label{barM-1}
     [\mathcal{J}_m,\mathcal{P}_n]=(m-n)\mathcal{P}_{m+n}+(\bar{\alpha}+\bar{\beta} m)\mathcal{M}_{m+n}.
\end{equation}
An interesting feature of the $\bar{M}(\bar{\alpha},\bar{\beta};\bar{\nu})$ algebra is that such symmetry is obtained by deforming the commutators $[\mathcal{J},\mathcal{J}]$ and $[\mathcal{J},\mathcal{P}]$ which are not the ideal part of the infinite dimensional algebra. As we known from the Hochschild-Serre factorization theorem, in the case of finite dimensional Lie algebra, the deformation of a Lie algebra can only be performed at the level of the ideal part without modifying the other commutators. Here, our result could confirm the conjecture made in \cite{Parsa:2018kys, Safari:2019zmc} in which the Hochschild-Serre factorization theorem might be extended for infinite dimensional algebras as follows: \textit{For infinite dimensional algebras with countable basis the deformations may appear in ideal and non-ideal parts, however, the deformations are always by coefficient in the ideal part.}

%%%%%%%%%%%%%%%%%%%%%%%%%%%%%%%%%%%%%%%%%%%%%%%%%%%%%%%%%%%%%%%%%%

\subsubsection{The \texorpdfstring{$M(a,b;c,d)$}{M(a,b;c,d)} algebra}
Another formal deformation is obtained by turning on simultaneously the functions $K$ and $\tilde{K}$. One can show from the Jacobi identities $[\mathcal{J},[\mathcal{J},\mathcal{P}]]+\text{cyclic permutations}=0$ and $[\mathcal{J},[\mathcal{J},\mathcal{M}]]+\text{cyclic permutations}=0$ that (see \eqref{firsteq-K} and \eqref{K-tildeK,O-tildeI})
\begin{equation}
K(m,n)=\alpha +\beta m, \qquad \tilde{K}(m,n)=2\alpha +2\beta m, \label{K-generic}
\end{equation}
which is only solution in all orders in functions.

The new algebra, which we name it as $M(a,b;c,d)$ algebra, has the following non vanishing commutation relations
\begin{equation} 
\begin{split}\label{new-Walgebra}
 & i[\mathcal{J}_m,\mathcal{J}_n]=(m-n)\mathcal{J}_{m+n}, \\
 &i[\mathcal{J}_m,\mathcal{P}_n]=-(bm+n+a)\mathcal{P}_{m+n},\\
 &i[\mathcal{J}_m,\mathcal{M}_n]=-(dm+n+c)\mathcal{M}_{m+n},\\
 &i[\mathcal{P}_m,\mathcal{P}_n]=(m-n)\mathcal{M}_{m+n},
\end{split}
\end{equation}
 where $c=2a=-2\alpha$ and $d=b-\beta=-2\beta-1$. One can show that such formal deformation can alternatively be obtained by considering the functions $K$ $\tilde{K}$ and $I$ simultaneously. Indeed, from the Jacobi identity $[\mathcal{J},[\mathcal{J},\mathcal{P}]]+\text{cyclic permutations}=0$ we have $K$ and $\tilde{K}$ are given by \eqref{K-generic} and $I=\xi(\alpha+\beta m)$. 
 Although such functions seems to induce a new formal deformation, one can use the same redefinition as in \eqref{redefine-I} to obtain
 \begin{equation}
     [\tilde{\mathcal{J}}_m,\tilde{\mathcal{P}}_n+F(n)\tilde{\mathcal{M}}_n]=(m-n+\alpha+\beta m)\left(\tilde{\mathcal{P}}_{m+n}+F(m+n)\tilde{\mathcal{M}}_{m+n}\right)+\xi(\alpha+\beta m)\tilde{\mathcal{M}}_{m+n},
\end{equation}
which reproduces the same algebra as \eqref{new-Walgebra} when $F(m)=constant=\xi$.

As it is discussed in \cite{Parsa:2018kys}, in context of $2d$ conformal field theory the parameters $b$ and $d$ are related to $h$ and $\tilde{h}$, which are the conformal weight of $\mathcal{P}$ and $\mathcal{M}$ respectively, through $b=1-h$ and $d=1-\tilde{h}$. On the other hand, the parameters $a$ (or $c$) is related to the periodicity properties of primary field $\mathcal{P}(\varphi)$ (or $\mathcal{M}(\varphi)$) through
\begin{equation*}
\mathcal{P}(\varphi+2\pi)=e^{2\pi i a}\mathcal{P}(\varphi),\qquad \mathcal{P}(\varphi)=\sum_n \mathcal{P}_n e^{i(n+a)\varphi}.
\end{equation*}
It is interesting to point out that diverse infinite dimensional structures appears for specific values of $a,b,c$ and $d$. In particular, let us suppose that $a=c=0$ in \eqref{new-Walgebra} and let us consider different values of $b,d$. First we set $b=0, d=1$ which leads to the algebra $M(0,0;0,1)$ with the following commutators
\begin{equation} 
\begin{split}\label{new-Walgebra-1}
 & i[\mathcal{J}_m,\mathcal{J}_n]=(m-n)\mathcal{J}_{m+n}, \\
 &i[\mathcal{J}_m,\mathcal{P}_n]=(-n)\mathcal{P}_{m+n},\\
 &i[\mathcal{J}_m,\mathcal{M}_n]=(-m-n)\mathcal{M}_{m+n},\\
 &i[\mathcal{P}_m,\mathcal{P}_n]=(m-n)\mathcal{M}_{m+n}.
\end{split}
\end{equation}
The generators $\mathcal{P}$ and $\mathcal{M}$ can be seen as a $U(1)$ current and a primary operator with conformal weight $h=0$,  respectively. The infinite dimensional algebra \eqref{new-Walgebra-1} corresponds to a Maxwellian version of the so-called $\mathfrak{u}(1)$ Kac-Moody algebra.
A different choice is $b=-\frac{1}{2},d=0$ which leads to a new algebra $M(0,-\frac{1}{2};0,0)$ whose non vanishing commutators are given by
\begin{equation} 
\begin{split}\label{twisted-virasoro-schrodinger}
 & i[\mathcal{J}_m,\mathcal{J}_n]=(m-n)\mathcal{J}_{m+n}, \\
 &i[\mathcal{J}_m,\mathcal{P}_n]=(\frac{m}{2}-n)\mathcal{P}_{m+n},\\
 &i[\mathcal{J}_m,\mathcal{M}_n]=(-n)\mathcal{M}_{m+n},\\
 &i[\mathcal{P}_m,\mathcal{P}_n]=(m-n)\mathcal{M}_{m+n},
\end{split}
\end{equation}
in which the generators $\mathcal{P}$ and $\mathcal{M}$ can be seen as a primary operator with conformal weight $h=\frac{3}{2}$ and a $U(1)$ current, respectively. This algebra is known as {\it twisted Schr\"{o}dinger-Virasoro algebra} \cite{Unterberger:2011yya}. In this reference the infinite enhancement of $3d$ Maxwell algebra, which is called $\mathfrak{sv}_{1}(0)$, is obtained as a deformation of the twisted Schr\"{o}dinger-Virasoro algebra.

When the indexes of the generator $\mathcal{P}$ are half integer valued the algebra corresponds to the so-called {\it Schr\"{o}dinger-Virasoro algebra} with spatial dimension $d=1$. The Schr\"{o}dinger-Virasoro algebra has a global part which is spanned by $6$ generators $\mathcal{J}_{0,\pm 1}$, $\mathcal{P}_{\pm\frac{1}{2}}$ and $\mathcal{M}_{0}$ where the latter appears as a central term. There are different works, for instance \cite{Alishahiha:2009nm, Compere:2009qm}, in which the authors have tried to find the Schr\"{o}dinger-Virasoro algebra as asymptotic symmetry of some spacetimes. 

An interesting feature of the $M(a,b;c,d)$ is that, as the $\bar{M}(\bar{\alpha},\bar{\beta};\bar{\nu})$ algebra, such deformation confirms the conjecture made in \cite{Parsa:2018kys, Safari:2019zmc}. Indeed, one can see that such deformation is obtained by considering coefficients in the ideal part.

Let us note that the family algebra $M(a,b;c,d)$, for some specific values of its parameters, can be deformed into new algebras out of this family. For example the \Max\ algebra given by $M(0,-1;0,-1)$  can be deformed in its ideal part into $\mathfrak{bms}_{3} \oplus \mathfrak{witt}$ as we shall see in the next section. Furthermore, the Schr\"{o}dinger-Virasoro algebra given by $M(0,\frac{1}{2};0,0)$ can be deformed in its $[\mathcal{J},\mathcal{J}]$ commutator. Despite this, it seems that the family algebra $M(a,b;c,d)$ is stable in the sense that for generic values of its parameters it can just be deformed into another family algebra $M(\bar{a},\bar{b};\bar{c},\bar{d})$ with shifted parameters. The latter should however be proved by direct computations.

As an ending remark, let us note that in
\cite{Compere:2009qm} they introduced the algebra with the same structure as $M(a,b;c,d)$. This algebra which is obtained with specific values of parameters as $M(\frac{z-2}{2z},\frac{-1}{z};\frac{z-2}{z},\frac{z-2}{z})$, is introduced as asymptotic symmetry algebra of Schr\"{o}dinger spacetimes.

%%%%%%%%%%%%%%%%%%%%%%%%%%%%%%%%%%%%%%%%%%%%%

\subsubsection{The \texorpdfstring{$\mathfrak{bms}_{3}\oplus\mathfrak{witt}$}{bms3+witt} algebra} 

A new formal deformation appears by studying the deformation of commutator $[\mathcal{P}_{m},\mathcal{P}_{m}]$ without modifying the other commutation relations. Indeed, as we have previously discussed, the Jacobi identity $[\mathcal{P},[\mathcal{P},\mathcal{P}]]$ $+\text{cyclic permutations}=0$ leads to relations \eqref{f1-O-tildeI} which is linear in functions. A non linear relation also appears from such Jacobi identity as
 \begin{equation}
     \begin{split}
         &(n-l)(m-n-l)f_{1}(n,l)f_{1}(m,l+n) + (l-m)(n-l-m)f_{1}(l,m)f_{1}(n,l+m)+\\
         &(m-n)(l-m-n)f_{1}(m,n)f_{1}(l,m+n) = 0. 
     \end{split}
 \end{equation}
with the same solution as $f_1(m,n)=\text{constant}$. The complete analysis to solve the equation for $f_{1}$ in the relation
\begin{equation}\label{f1h1}
(n-l)(m-n-l)f_{1}(m,l+n)+ (l-m)(n-l-m)f_{1}(n,l+m)+(m-n)(l-m-n)f_{1}(m,n)=0,
\end{equation}
can be found in \cite{Parsa:2018kys}.
One can show that the new algebra 
\begin{equation} 
\begin{split}\label{f1-deform-maxwell}
 & i[\mathcal{J}_m,\mathcal{J}_n]=(m-n)\mathcal{J}_{m+n}, \\
 &i[\mathcal{J}_m,\mathcal{P}_n]=(m-n)\mathcal{P}_{m+n},\\
 &i[\mathcal{J}_m,\mathcal{M}_n]=(m-n)\mathcal{M}_{m+n},\\
 &i[\mathcal{P}_m,\mathcal{P}_n]=(m-n)\mathcal{M}_{m+n}+\varepsilon(m-n)\mathcal{P}_{m+n}.\\
\end{split}
\end{equation}
 obtained by $f_{1}(m,n)=\varepsilon$, with $\varepsilon$ being an arbitrary constant, is not isomorphic to the original algebra and hence the deformation is non trivial. By a redefinition of generators \footnote{The deformation parameter can be removed by an appropriate redefinition as $\mathcal{P}\equiv\varepsilon\mathcal{P}$ and $\mathcal{M}\equiv\varepsilon^{2}\mathcal{M}$. } as \begin{equation} 
\begin{split}\label{redefine-pp}
  \mathcal{J}_m\equiv L_{m}+S_{m}, \qquad &\mathcal{P}_m\equiv T_{m}+S_{m}, \qquad \mathcal{M}_m\equiv - T_{m},
\end{split}
\end{equation}
one reaches to the new algebra with non vanishing commutators
\begin{equation} 
\begin{split}\label{bms3+witt-2}
 & i[{L}_m,{L}_n]=(m-n){L}_{m+n}, \\
 &i[{L}_m,{T}_n]=(m-n){T}_{m+n},\\
 &i[{S}_m,{S}_n]=(m-n){S}_{m+n}.
\end{split}
\end{equation}
The new algebra \eqref{bms3+witt-2} has the direct sum structure as $\mathfrak{bms}_{3}\oplus\mathfrak{witt}$. 
The global part of the algebra \eqref{bms3+witt-2} corresponds to the $\mathfrak{iso}(2,1)\oplus\mathfrak{so}(2,1)$ algebra when we restrict ourselves to $m,n=\pm1,0$ which is the direct sum of the $3d$ Poincar\'{e}  and the $3d$ Lorentz algebras. Such finite structure has also been obtained as a deformation of the $d=2+1$ Maxwell algebra in \cite{Gomis:2009dm} but not at the same basis as \eqref{f1-deform-maxwell}. Note also that this algebra is a subalgebra of $W(0,-1;0,0)$, which is obtained as deformation of $\mathfrak{bms}_{4}$ algebra \cite{Safari:2019zmc}.

Interestingly the same structure can be obtained by turning on $f_{1}$ and $g_{2}$ simultaneously. In fact we have from the Jacobi identity $[\mathcal{P},[\mathcal{P},\mathcal{M}]]+\text{cyclic permutations}=0$ the following relation
\begin{equation}\label{g2f1-formal}
g_{2}(n,l)g_{2}(m,n+l)-g_{2}(m,l)g_{2}(n,l+m)-(m-n)g_{2}(m+n,l)f_{1}(m,n)=0,
\end{equation}
which is solved for $g_{2}(m,n)=\varepsilon(m-n)$ and $f_{1}(m,n)=\varepsilon$. Let us note that $g_{2}(m,n)=\varepsilon(m-n)$ comes directly from a relation similar to \eqref{f,g,h-2(1)} as a consequence of the Jacobi identity $[\mathcal{P},[\mathcal{M},\mathcal{J}]]+\text{cyclic permutations}=0$. So the commutators of the new algebra obtained through this deformation are
\begin{equation} 
\begin{split}\label{f1g2-deform-maxwell}
 & i[\mathcal{J}_m,\mathcal{J}_n]=(m-n)\mathcal{J}_{m+n}, \\
 &i[\mathcal{J}_m,\mathcal{P}_n]=(m-n)\mathcal{P}_{m+n},\\
 &i[\mathcal{J}_m,\mathcal{M}_n]=(m-n)\mathcal{M}_{m+n},\\
 &i[\mathcal{P}_m,\mathcal{P}_n]=(m-n)\mathcal{M}_{m+n}+\varepsilon(m-n)\mathcal{P}_{m+n},\\
 &i[\mathcal{P}_m,\mathcal{M}_n]=\varepsilon(m-n)\mathcal{M}_{m+n},\\
 &i[\mathcal{M}_m,\mathcal{M}_n]=0. 
\end{split}
\end{equation}
One can show that the $\mathfrak{bms}_{3}\oplus\mathfrak{witt}$ algebra appears by considering an appropriate redefinition of the generators as\footnote{For convenience we drop the deformation parameter in our redefinition since it can be absorbed by an appropriate redefinition of the generators.}
\begin{equation} 
\begin{split}\label{redefine-f1g2}
  \mathcal{J}_m\equiv L_{m}+S_{m}, \qquad \mathcal{P}_m\equiv L_{m}+T_{m}, \qquad \mathcal{M}_m\equiv T_{m}.
\end{split}
\end{equation}

\subsubsection{The \texorpdfstring{$\mathfrak{witt}\oplus\mathfrak{witt}\oplus\mathfrak{witt}$}{witt+witt+witt} algebra}

Three copies of the Witt algebra can be obtained through deformations induced by two or more functions simultaneously and after an appropriate redefinition of the generators. Here, based on \eqref{f2-tildeO} for $f_2$ and using \eqref{f2-g3-h1} for $g_{3}$, we shall turn on two functions simultaneously. Indeed, the Jacobi identity $[\mathcal{M},[\mathcal{M},\mathcal{P}]]+\text{cyclic permutations}=0$ and $[\mathcal{M},[\mathcal{M},\mathcal{M}]]+\text{cyclic permutations}=0$ gives rise to non linear relations as
\begin{equation}\label{f2-g3-formal}
 f_{2}(l,n)f_{2}(l+n,m)- f_{2}(l,m)f_{2}(l+m,n)+(m-n)g_{3}(m,n)f_{2}(l,m+n)=0,
\end{equation}
and 
\begin{multline}\label{g3-formal}
 (n-l)(m-n-l)g_{3}(n,l)g_{3}(m,l+n)+ (l-m)(n-l-m)g_{3}(l,m)g_{3}(n,l+m)+\\ +
  (m-n)(l-m-n)g_{3}(m,n)g_{3}(l,m+n)=0,
\end{multline}
where is also satisfied with the solutions $f_2(m,n)=\lambda (m-n)$ and $g_3(m,n)=\lambda$. Then we find the following formal deformation of the \Max\ algebra
 \begin{equation} 
\begin{split}\label{PM-deform-maxwell4}
 & i[\mathcal{J}_m,\mathcal{J}_n]=(m-n)\mathcal{J}_{m+n}, \\
 &i[\mathcal{J}_m,\mathcal{P}_n]=(m-n)\mathcal{P}_{m+n},\\
 &i[\mathcal{J}_m,\mathcal{M}_n]=(m-n)\mathcal{M}_{m+n},\\
 &i[\mathcal{P}_m,\mathcal{P}_n]=(m-n)\mathcal{M}_{m+n},\\
 &i[\mathcal{P}_m,\mathcal{M}_n]=\lambda(m-n)\mathcal{P}_{m+n},\\
 &i[\mathcal{M}_m,\mathcal{M}_n]=\lambda(m-n)\mathcal{M}_{m+n}. 
\end{split}
\end{equation}
Upon the following redefinition of the generators,
\begin{equation} 
\begin{split}\label{redefine-f2g3}
  \mathcal{J}_m\equiv L_{m}+T_{m}+S_{m}, \qquad \mathcal{P}_m\equiv L_{m}-T_{m}, \qquad \mathcal{M}_m\equiv L_{m}+T_{m}.
\end{split}
\end{equation}
the above algebra reproduces three copies of the Witt algebra
\begin{equation} 
\begin{split}\label{3witt}
 & i[L_m,L_n]=(m-n)L_{m+n}, \\
 &i[T_m,T_n]=(m-n)T_{m+n},\\
 &i[S_m,S_n]=(m-n)S_{m+n}.
\end{split}
\end{equation}
This result is the infinite dimensional generalization of the one obtained in \cite{Gomis:2009dm} for the $2+1$ Maxwell algebra which was called $k-$deformation. In particular, they showed that the $k-$deformation leads to one of $\mathfrak{so}(2,2)\oplus\mathfrak{so}(2,1)$ or $\mathfrak{so}(3,1)\oplus\mathfrak{so}(2,1)$ algebras depending on the sign of the deformation parameter. On the other hand, the three copies of the Witt algebra have three $\mathfrak{sl}(2,\mathbb{R})$ algebras as their global part. In this specific basis both $\mathfrak{so}(2,2)$ and $\mathfrak{so}(3,1)$ are written as $\mathfrak{sl}(2,\mathbb{R})\oplus\mathfrak{sl}(2,\mathbb{R})$, while $\mathfrak{so}(2,1)$ is written as $\mathfrak{sl}(2,\mathbb{R})$. At the gravity level, the so-called AdS-Lorentz algebra, which can be written as three $\mathfrak{so}(2,1)$, allows to accommodate a cosmological constant to the three-dimensional Maxwell Chern-Simons gravity action \cite{Hoseinzadeh:2014bla, Diaz:2012zza, Concha:2018jjj}. 

It is interesting to note that three copies of the Witt algebra can alternatively obtained by turning on other functions. Indeed one can easily verify that $f_1(m,n)=g_3(m,n)=\delta$ and $f_2=\delta(m-n)$ also reproduces such structure. The formal deformations induced by two functions simultaneously as $h_1$ and $g_3$ or $h_2$ and $f_3$ also reproduce the three copies of the Witt algebra after an appropriate redefinition of the generators. It is important to clarify that such deformations with coefficients being not in the ideal part can be obtained as a redefinition of a deformed \Max\ algebra with coefficients in the ideal part such that the conjecture presented in \cite{Parsa:2018kys, Safari:2019zmc} about a possible extension of the Hochschild-Serre factorization theorem is still valid.

One could conjecture that, based on the analysis done for the direct sum of two Witt algebras \cite{Parsa:2018kys}, the direct sum of three Witt algebra is rigid. Furthermore, one could expect to recover the $\mathfrak{witt}\oplus\mathfrak{witt}\oplus\mathfrak{witt}$ algebra as a deformation of the $\mathfrak{bms}_3\oplus\mathfrak{witt}$ algebra since we know that the $\mathfrak{bms}_3$ algebra is not stable and can be deformed to two copies of the Witt algebra.

\subsection{Algebraic cohomology argument}\label{sec:cohomology}

Until now we have classified all possible nontrivial infinitesimal and formal deformations of the \Max\ algebra by studying the Jacobi identities. As discussed in \cite{Parsa:2018kys}, one can approach and analyze such issue by cohomology consideration. Indeed one can classify all infinitesimal deformations of the \Max\ algebra by computing $\mathcal{H}^{2}(\mathfrak{Max}_{3};\mathfrak{Max}_{3})$. In our previous works, in which we tackled Lie algebras with abelian ideal, we used the theorem 2.1 of \cite{degrijse2009cohomology} which is  crucial for cohomological consideration. Nonetheless, we cannot use this theorem here since \Max\ does not have abelian ideal. We shall only state our result in cohomological language. As we can see from the our results in previous part, we have just four formal deformations for the \Max\ algebra. It is obvious that both $M(a,b;c,d)$ and $\bar{M}(\bar{\alpha},\bar{\beta};\bar{\nu})$ family algebras are deformed by the $K, \tilde{K}, I$ and $F$ terms, with coefficients from ideal part, $\mathcal{P}$ and $\mathcal{M}$. The same argument is true for the new algebra  $\mathfrak{bms}_{3} \oplus \mathfrak{witt}$ which is obtained through deformation induced by $f_{1}$ with coefficient in $\mathcal{P}$. The three copies of the Witt algebra can be obtained via deformation induced by $h_{1}, g_{3}$ or $h_{2}, f_{3}$  and also by $f_{2},g_{3}$, which means that the two first cases are just a redefinition of the latter. As summary, we have shown that, unlike the Hochschild-Serre factorization theorem of finite Lie algebras, other commutators of \Max\ algebra, except the ideal part, can also be deformed but only by terms with coefficients from the ideal part. As it has been discussed in the works \cite{Parsa:2018kys, Safari:2019zmc} this result can be viewed as an extension of the Hochschild-Serre factorization theorem for infinite dimensional algebras. \footnote{Here we are tackling infinite dimensional Lie algebras which are extensions of the Witt algebra.} 
 
 In the cohomological language our results for the \Max\ algebra can be written as
\begin{equation}
    \mathcal{H}^{2}(\mathfrak{Max}_{3};\mathfrak{Max}_{3})\cong \mathcal{H}^{2}(\mathfrak{Max}_{3};\mathfrak{h}).
\end{equation}
where $\mathfrak{h}$ denotes the ideal part of \Max\ algebra spanned by generators $\mathcal{P}$ and $\mathcal{M}$.

\section{Central extensions of the deformed \texorpdfstring{$\mathfrak{Max}_{3}$}{Max3} algebras}\label{sec:4}

In this section, we present explicit central extensions of the infinite-dimensional algebras obtained as a deformation of the $\mathfrak{Max}_3$ algebra introduced previously. In particular, one of the central extension reproduces a known asymptotic symmetry of a three-dimensional gravity theory.

\subsection{Central extension of the \texorpdfstring{$\mathfrak{bms}_{3} \oplus \mathfrak{witt}$}{bms3+witt} and the \texorpdfstring{$\mathfrak{witt}\oplus\mathfrak{witt}\oplus\mathfrak{witt}$}{witt+witt+witt} algebra}

We have shown that ones of the deformations of the \Max\ algebra are given by the $\mathfrak{bms}_{3} \oplus \mathfrak{witt}$ and three copies of the Witt algebra. In this section we briefly review the known central extensions of the $\mathfrak{bms}_3$ and the Witt algebra.

The most general central extension of the $\mathfrak{bms}_{3} \oplus \mathfrak{witt}$ is given by
\begin{equation} 
\begin{split}\label{bms3+witt_CE}
 & i[{L}_m,{L}_n]=(m-n){L}_{m+n}+\frac{c_{LL}}{12}m^{3}\delta_{m+n,0}, \\
 &i[{L}_m,{T}_n]=(m-n){T}_{m+n}+\frac{c_{LT}}{12}m^{3}\delta_{m+n,0},\\
 &i[{S}_m,{S}_n]=(m-n){S}_{m+n}+\frac{c_{SS}}{12}m^{3}\delta_{m+n,0},
\end{split}
\end{equation}
where the central charges $c_{LL}$, $c_{LT}$ and $c_{SS}$ can be related to three independent terms of the Chern-Simons $\mathfrak{iso}(2,1)\oplus \mathfrak{so}(2,1)$ gravity action as follows:
\begin{equation}
\begin{split}\label{CentralTerms4}
c_{LL}=12k\alpha_{0}, \qquad c_{LT}=12k\alpha_{1}, \qquad c_{SS}=12k\beta_{0},
\end{split}
\end{equation}
where $\alpha_{0}$ and $\alpha_{1}$ are the respective coupling constants appearing in the three-dimensional Chern-Simons Poincar\'{e} gravity. On the other hand, $\beta_{0}$ is the coupling constant of the exotic Lagrangian invariant under the $\mathfrak{so}(2,1)$ algebra.  It would be interesting to explore the central terms in the basis $\{\mathcal{J}_{m},\mathcal{P}_{m},\mathcal{M}_{m}\}$ and the possibility that the central extensions of the infinite-dimensional algebras \eqref{f1-deform-maxwell} and \eqref{f1g2-deform-maxwell} appears as the asymptotic symmetries of three-dimensional gravity theory invariant under deformations of the Maxwell algebra.

On the other hand, a central extension for the $\mathfrak{witt}\oplus\mathfrak{witt}\oplus\mathfrak{witt}$ algebra is naturally given by\begin{equation} 
\begin{split}\label{3Vir}
 & i[L_m,L_n]=(m-n)L_{m+n}+\frac{c_{LL}}{12}m^{3}\delta_{m+n,0}, \\
 &i[T_m,T_n]=(m-n)T_{m+n}+\frac{c_{TT}}{12}m^{3}\delta_{m+n,0},\\
 &i[S_m,S_n]=(m-n)S_{m+n}+\frac{c_{SS}}{12}m^{3}\delta_{m+n,0}.
\end{split}
\end{equation}
Interestingly, considering the following redefinition of the generators
\begin{equation} 
\begin{split}\label{redefine-AdSL}
L_m\equiv \frac{1}{2}(\mathcal{M}_{m}+\mathcal{P}_{m}),\quad T_m\equiv \frac{1}{2}(\mathcal{M}_{m}-\mathcal{P}_{m}), \quad S_m\equiv \mathcal{J}_{m}-\mathcal{M}_{m},
\end{split}
\end{equation}
 and the following redefinition of the central terms
\begin{equation} 
\begin{split}\label{redefine-centralterms}
 c_{LL}\equiv \frac{1}{2}(c_{JM}+c_{JP}), \quad c_{TT}\equiv \frac{1}{2}(c_{JM}-c_{JP}), \quad c_{SS}\equiv (c_{JJ}-c_{JM}).
\end{split}
\end{equation}
we recover the asymptotic symmetry of the Chern-Simons gravity theory invariant under the so-called AdS-Lorentz algebra \cite{Concha:2018jjj}. Such symmetry has been previously studied in \cite{Soroka:2006aj, Gomis:2009dm, Hoseinzadeh:2014bla, Diaz:2012zza} and extended to higher dimensions in \cite{Concha:2016kdz, Concha:2016tms, Concha:2017nca} in Lovelock theory.

\subsection{Central extension of \texorpdfstring{$M(a,b;c,d)$}{M(a,b;c,d)}}
Here we shall classify the central terms of the $M(a,b;c,d)$ algebra. One can easily find that the $M(a,b;c,d)$ algebra for generic values of parameter space $a,b,c$ and $d$ admits only one central term in its Witt subalgebra. However there are some specific points in which it is possible to have other non-trivial central terms. We follow the results of the work \cite{gao2011low} which classifies the central terms of $W(a,b)$ algebra.

\subsubsection{Central terms for specific points in parameters space of \texorpdfstring{$M(a,b;c,d)$}{M(a,b;c,d)}}
\paragraph{ \texorpdfstring{$M(0,0;0,1)$}{M(d=1,a,b,c=0)} case.}
By setting the parameters as $a,b,c=0,d=1$ we obtain a new algebra with non vanishing commutators as in  \eqref{new-Walgebra-1}. One can readily check that there is a central term in the Witt subalgebra given by $c_{JJ}m^{3}$ %of \eqref{beta=-1}% 
so we shall take it in account in what follows.
Let us consider now the central term as $[\mathcal{J}_m,\mathcal{P}_n]=(-n)\mathcal{P}_{m+n}+S(m,n)$ where $S(m,n)$ is an arbitrary function. One can see that the Jacobi identity $[\mathcal{J},[\mathcal{J},\mathcal{P}]]+\text{cyclic permutations}=0$ implies the following constraint
\begin{equation}\label{eq-S}
   -lS(m,n+l)+lS(n,l+m)+(n-m)S(m+n,l)=0,
\end{equation}
If the function $S(m,n)$ is a symmetric function by setting $l=0$ one obtains $S(m+n,0)=0$. Then the only solution is $S(m,n)=c_{JP}m^{2}\delta_{m+n,0}$ in which $c_{JP}$ is an arbitrary constant as expected from central extension of the $u(1)$ Kac-Moody algebra \cite{Parsa:2018kys}. On the other hand, one can show that there is no solution for $S(m,n)$ being an anti symmetric function. The rest of the Jacobi identities do not put additional constraints on $S(m,n)$ reproducing a non trivial central extension. %so one concludes that this is a non trivial central extension. 
Another central term can appear as $[\mathcal{J}_m,\mathcal{M}_n]=(-m-n)\mathcal{M}_{m+n}+T(m,n)$ where $T(m,n)$ is an arbitrary function. The Jacobi identity $[\mathcal{J},[\mathcal{J},\mathcal{M}]]+\text{cyclic permutations}=0$ leads to
\begin{equation}\label{eq-T}
    (-n-l)T(m,n+l)+(m+l)T(n,l+m)+(n-m)T(m+n,l)=0.
\end{equation}
If the function $T(m,n)$ is a symmetric function one obtains $T(m,n)=T(m+n,0)=\bar{T}(m+n)$. Then we have $T(m,n)=(c_{JM1}m+c_{JM2})\delta_{m+n,0}$ where $c_{JM1,2}$ are arbitrary constants. On the other hand the Jacobi identity $[\mathcal{P},[\mathcal{P},\mathcal{J}]]+\text{cyclic permutations}=0$ implies $T(m,n)=0$. One can also see that there is no solution for $T(m,n)$ being an anti symmetric function. 
Let us consider now the presence of central terms in both $[\mathcal{J}_m,\mathcal{M}_n]=(-m-n)\mathcal{M}_{m+n}+T(m,n)\delta_{m+n+l,0}$ and $[\mathcal{P}_m,\mathcal{P}_n]=(m-n)\mathcal{M}_{m+n}+U(m,n)\delta_{m+n,0}$ simultaneously. The Jacobi identity  $[\mathcal{P},[\mathcal{P},\mathcal{J}]]+\text{cyclic permutations}=0$ leads to
\begin{equation}\label{eq-T-U}
    \left((n)U(m,n+l)-(m)U(n,l+m)+(m-n)T(l,m+n)\right)\delta_{m+n+l,0}=0,
\end{equation}
which does not have a non zero solution for $U(m,n)$ when $T(m,n)=c_{JM1}m$. However when we consider $T(m,n)=c_{JM2}$, one finds $U(m,n)=c_{JM2}$ which represents another non trivial central extension. An additional central term can appear in $[\mathcal{P}_m,\mathcal{P}_n]=(m-n)\mathcal{M}_{m+n}+U(m,n)\delta_{m+n,0}$ when other central terms are turned off. The Jacobi identities $[\mathcal{P},[\mathcal{P},\mathcal{P}]]+\text{cyclic permutations}=0$ and $[\mathcal{P},[\mathcal{P},\mathcal{M}]]+\text{cyclic permutations}=0$ do not constrain $U(m,n)$. The only remaining Jacobi identity is $[\mathcal{P},[\mathcal{P},\mathcal{J}]]+\text{cyclic permutations}=0$ which implies
\begin{equation}\label{eq-U}
    \left((n)U(m,n+l)-(m)U(n,l+m)\right)\delta_{m+n+l,0}=0,
\end{equation}
with $U(m,n)=c_{PP}m$. However, it is possible to see that the following redefinition 
\begin{equation}
    \mathcal{M}_{m}\equiv \tilde{\mathcal{M}}_{m}+c\delta_{m,0},
\end{equation}
do not reproduce a non trivial central extension for $c=-\frac{c_{PP}}{2}$ since the central term $c_{PP}$ can be absorbed.

To summarize, the most general central extension of $M(0,0;0,1)$ is
%\pnote{I think here it should be $M(0,0)$ or $M(0,0;0,-1)$ depending on your preferred notation.}
\begin{equation} 
\begin{split}\label{beta=-1}
 & i[\mathcal{J}_m,\mathcal{J}_n]=(m-n)\mathcal{J}_{m+n}+\frac{c_{JJ}}{12}m^{3}\delta_{m+n,0}, \\
 &i[\mathcal{J}_m,\mathcal{P}_n]=(-n)\mathcal{P}_{m+n}+c_{JP}m^{2}\delta_{m+n,0},\\
 &i[\mathcal{J}_m,\mathcal{M}_n]=(-m-n)\mathcal{M}_{m+n}+c_{JM}\delta_{m+n,0},\\
 &i[\mathcal{P}_m,\mathcal{P}_n]=(m-n)\mathcal{M}_{m+n}+c_{JM}\delta_{m+n,0}.
\end{split}
\end{equation}

\paragraph{\texorpdfstring{$M(0,-2;0,-3)$}{a=c=0,b=-2,d=-3} case.}
The next values of the parameters which we will consider is $a=c=0,b=-2,d=-3$ for which we obtain a new algebra with non vanishing commutators as
\begin{equation} 
\begin{split}\label{beta=1,1}
 & i[\mathcal{J}_m,\mathcal{J}_n]=(m-n)\mathcal{J}_{m+n}, \\
 &i[\mathcal{J}_m,\mathcal{P}_n]=(-m-n)\mathcal{P}_{m+n},\\
 &i[\mathcal{J}_m,\mathcal{M}_n]=(-3m-n)\mathcal{M}_{m+n},\\
 &i[\mathcal{P}_m,\mathcal{P}_n]=(m-n)\mathcal{M}_{m+n}.
\end{split}
\end{equation}
Let us consider first the central term in $[\mathcal{J}_m,\mathcal{P}_n]=(-m-n)\mathcal{P}_{m+n}+S(m,n)$. The Jacobi identity $[\mathcal{J},[\mathcal{J},\mathcal{P}]]+\text{cyclic permutations}=0$ reproduces the same constraint as \eqref{eq-T} on $S(m,n)$. So we obtain $S(m,n)=(c_{JP1}m+c_{JP2})\delta_{m+n,0}$. One can turn on a central term as $[\mathcal{J}_m,\mathcal{M}_n]=(-3m-n)\mathcal{M}_{m+n}+T(m,n)$. The Jacobi identity $[\mathcal{M},[\mathcal{M},\mathcal{J}]]+\text{cyclic permutations}=0$ implies
\begin{equation}
   -(3n+l)S(m,l+n)+(3m+l)S(n,l+m)+(n-m)S(m+n,l)=0,
\end{equation}
which has no non trivial solution leading to $T(m,n)=0$. On the other hand one may consider the central term as $[\mathcal{P}_m,\mathcal{P}_n]=(m-n)\mathcal{M}_{m+n}+U(m,n)\delta_{m+n,0}$ however this does not lead to a non trivial central term.  
Therefore, there is no further central extensions for $M(a=c=0,b=-2,d=-3)$ and the most general central extension of this %$M(0,-2;0,-3)$ 
algebra is given by
\begin{equation} 
\begin{split}\label{beta=1}
 & i[\mathcal{J}_m,\mathcal{J}_n]=(m-n)\mathcal{J}_{m+n}+\frac{c_{JJ}}{12}m^{3}\delta_{m+n,0}, \\
 &i[\mathcal{J}_m,\mathcal{P}_n]=(-m-n)\mathcal{P}_{m+n}+(c_{JP1}m+c_{JP2})\delta_{m+n,0},\\
 &i[\mathcal{J}_m,\mathcal{M}_n]=(-3m-n)\mathcal{M}_{m+n},\\
 &i[\mathcal{P}_m,\mathcal{P}_n]=(m-n)\mathcal{M}_{m+n}.
\end{split}
\end{equation}
As we can see this is in contradiction with the result of theorem 5.7. of \cite{Roger:2006rz} in which they did not mention the term $c_{JP1}\delta_{m+n,0}$ in \eqref{beta=1}.

\paragraph{\texorpdfstring{$M(0,-\frac{1}{2};0,0)$}{beta=-1,alpha=0} case.}
Another value of the parameters that one could explore is $a=c=0,b=-\frac{1}{2},d=0$ which leads to the new algebra \eqref{twisted-virasoro-schrodinger}. As  mentioned before this algebra is known as the twisted Schr\"{o}dinger-Virasoro algebra. According to the theorem 2.2 in \cite{gao2011low} we know that there is no central term in the $[\mathcal{J}_m,\mathcal{P}_n]$ commutator.\footnote{This can be easily checked by adding a central term like $S(m,n)$ to this commutator and considering the Jacobi identity $[\mathcal{J},[\mathcal{J},\mathcal{P}]]+\text{cyclic permutations}=0$.} One can indeed show from the Jacobi identity that the only central extension for the twisted Schrödinger-Virasoro algebra appears in its Witt subalgebra:
\begin{equation} 
\begin{split}\label{beta=-1/2}
 & i[\mathcal{J}_m,\mathcal{J}_n]=(m-n)\mathcal{J}_{m+n}+\frac{c_{JJ}}{12}m^{3}\delta_{m+n,0}, \\
 &i[\mathcal{J}_m,\mathcal{P}_n]=(\frac{m}{2}-n)\mathcal{P}_{m+n},\\
 &i[\mathcal{J}_m,\mathcal{M}_n]=(-n)\mathcal{M}_{m+n},\\
 &i[\mathcal{P}_m,\mathcal{P}_n]=(m-n)\mathcal{M}_{m+n}.
\end{split}
\end{equation} %Then we consider the central term %$[\mathcal{J}_m,\mathcal{M}_n]=(-n)\mathcal{M}_{m+n}+T(m,n)$. Although the Jacobi identity $[\mathcal{J},[\mathcal{J},\mathcal{M}]]+\text{cyclic permutations}=0$ leads to a relation similar to \eqref{eq-S} which implies $T(m,n)=(c_{JM}m^{2})\delta_{m+n,0}$, one can see that the Jacobi $[\mathcal{P},[\mathcal{P},\mathcal{J}]]+\text{cyclic permutations}=0$ yields $T(m,n)=0$. One can check the possibility of  simultaneous central terms $[\mathcal{J}_m,\mathcal{M}_n]=(-n)\mathcal{M}_{m+n}+T(m,n)$ and $[\mathcal{P}_m,\mathcal{P}_n]=(m-n)\mathcal{M}_{m+n}+U(m,n)\delta_{m+n,0}$. The Jacobi identity $[\mathcal{P},[\mathcal{P},\mathcal{J}]]+\text{cyclic permutations}=0$ leads to 
%\begin{equation}\label{T-U-bet-1/2}
 %   \left((n-l)U(m,n+l)+(l-m)U(n,l+m)\right)\delta_{m+n+l,0}+(m-n)T(l,m+n)=0.
%\end{equation}
%Replacing the solution $T(m,n)=(c_{JM}m^{2})\delta_{m+n,0}$ into \eqref{T-U-bet-1/2}, one finds  $U(m,n)=0$.
%Then one can see that there is no non zero solution for $T(m,n)$ and $U(m,n)$. 
%One can also check that the Jacobi identity 
%$[\mathcal{P},[\mathcal{P},\mathcal{M}]]+\text{cyclic permutations}=0$ does not allow  addition of a central term in the commutator $[\mathcal{M},\mathcal{M}]$. We conclude that the only central extension for $M(0,-\frac{1}{2};0,0)$  (twisted Schr\"{o}dinger-Virasoro algebra) appears in its Witt subalgebra part

%\pnote{I think it should be $M(0,\frac{1}{2})$ or $M(0,\frac{1}{2};0,0)$}
%%%%%%%%%%%%%

%\hnote{I considered central extensions of $W(a,0)$ algebra which is appeared as deformation of \bms . I think that there is a central extension $[\mathcal{P}_m,\mathcal{P}_n]=f(m,n)\delta_{m+n,0}$ for all values of $a$ where Jacobi $[\mathcal{J}_{m},[\mathcal{P}_{n},\mathcal{P}_{l}]]+\text{cyclic permutations}=0$ leads to $f(m,n)=m+a$. Am I right? So I think the result of relation 2.7 in \cite{gao2011low} should be corrected! }

\subsection{Central extension of \texorpdfstring{$\bar{M}(\bar{\alpha},\bar{\beta};\bar{\nu})$}{M(bar-alpha,bar-beta)}}

As we have mentioned the functions $I(m,n)$ and $F(m,n)$ are just constrained by the Jacobi identities $[\mathcal{J},[\mathcal{J},\mathcal{J}]]+\text{cyclic permutations}=0$ and $[\mathcal{J},[\mathcal{J},\mathcal{P}]]+\text{cyclic permutations}=0$. Let us then  consider the central terms constrained by these Jacobi identities. In particular, let us first consider the central term as 
$[\mathcal{J}_m,\mathcal{J}_n]=(m-n)\mathcal{J}_{m+n}+\bar{\nu}(m-n)\mathcal{P}_{m+n}+R(m,n)\delta_{m+n,0}$. From the Jacobi identity $[\mathcal{J},[\mathcal{J},\mathcal{J}]]+\text{cyclic permutations}=0$ we find the solution $R(m,n)=c_{JJ}m^{3}$. Let $S(m,n)$ be an arbitrary functions which appears in $[\mathcal{J}_m,\mathcal{P}_n]=(m-n)\mathcal{P}_{m+n}+(\bar{\alpha}+\bar{\beta}m)\mathcal{M}_{m+n}+S(m,n)$ and satisfy the following constraint
\begin{equation}
   (n-l)S(m,l+n)+(l-m)S(n,l+m)+(n-m)S(m+n,l)=0.
\end{equation}
The Jacobi identities $[\mathcal{J},[\mathcal{J},\mathcal{J}]]+\text{cyclic permutations}=0$ and $[\mathcal{J},[\mathcal{J},\mathcal{P}]]+\text{cyclic permutations}=0$, as expected, indicate the existence of a central term $S(m,n)=c_{JP}m^{3}\delta_{m+n,0}$. One can see that a central term can also appear in the commutator $[\mathcal{J}_m,\mathcal{M}_n]=(m-n)\mathcal{M}_{m+n}+T(m,n)$ where $T(m,n)$ is an arbitrary function. From the Jacobi identity $[\mathcal{J},[\mathcal{J},\mathcal{M}]]+\text{cyclic permutations}=0$ we find that the function is fixed as $T(m,n)=c_{JM}m^{3}\delta_{m+n,0}$ if we also turn on the same central term in $[\mathcal{P}_m,\mathcal{P}_n]=(m-n)\mathcal{M}_{m+n}+U(m,n)$ with $U(m,n)=c_{JM}\delta_{m+n,0}$. However one should also consider the Jacobi identity $[\mathcal{J},[\mathcal{J},\mathcal{P}]]+\text{cyclic permutations}=0$ which leads to
\begin{equation}
  c_{JM} \left((\bar{\alpha}+\bar{\beta}n-\bar{\nu}l)m^{3}-(\bar{\alpha}+\bar{\beta}m+\bar{\nu}l)n^{3}+\bar{\nu}(m-n)l^{3}\right)\delta_{m+n+l,0}=0.
\end{equation}
Let us note that since the three parameters $\bar{\alpha}, \bar{\beta}$ and $\bar{\nu}$ are independent, there is no solution for the above expression for $\bar{\alpha},\bar{\beta},\bar{\nu}\neq 0$. Nevertheless for $\bar{\alpha}=\bar{\nu}=0$, we have the non trivial central extension $T(m,n)=U(m,n)=c_{JM}m^{3}\delta_{m+n,0}$. Thus, we conclude that the most general central extension for the $\bar{M}(0,\bar{\beta};0)$ algebra is given by
\begin{equation} 
\begin{split}
 & i[\mathcal{J}_m,\mathcal{J}_n]=(m-n)\mathcal{J}_{m+n}+\frac{c_{JJ}}{12}m^{3}\delta_{m+n,0}, \\
 &i[\mathcal{J}_m,\mathcal{P}_n]=(m-n)\mathcal{P}_{m+n}+\bar{\beta}m\mathcal{M}_{m+n}+\frac{c_{JP}}{12}m^{3}\delta_{m+n,0},\\
 &i[\mathcal{J}_m,\mathcal{M}_n]=(-n)\mathcal{M}_{m+n}+\frac{c_{JM}}{12}m^{3}\delta_{m+n,0},\\
 &i[\mathcal{P}_m,\mathcal{P}_n]=(m-n)\mathcal{M}_{m+n}+\frac{c_{JM}}{12}m^{3}\delta_{m+n,0}.
\end{split}
\end{equation}

%%%%%%%%%%%%%%%%%%%%%%%%%%%%%%%%%%%%%%%%%%%%%%%%%%%%%%%%%%%%%%%%%%%%%%%%%%%%%%%%%%%%%%%%%%%%%%%%%%%%%%%%

%--------------------------------------------------------------------------

%----------------------------------------------------------------------------------------------

 %-------------------------------------------------------------------------------------------------------------------------------------------------

%-------------------------------------------------------------------------------------------------------------------------------------------------------------

%%%%%%%%%%%%%%%%%%%%%%%%%%%%%%%%%%%%%%%%%%%%%%%%%%%%%%%%%%%%%%%%%%%%%%%%%%%%%%%%%%%%%%%%%%%%%%%%%%%%%%%%%%

\section{Summary and concluding remarks}\label{sec-discussions}
In this work we have considered the deformation and stability of \Max\ algebra which is the infinite enhancement of the $2+1$ dimensional Maxwell algebra and describes the asymptotic symmetry of the Chern-Simons gravity theory invariant under the Maxwell algebra \cite{Concha:2018zeb}. We have shown that the \Max\ algebra is not stable and can be deformed to four possible formal deformations. The \Max\ algebra can be formally deformed into $\mathfrak{bms}_{3}\oplus\mathfrak{witt}$ or three copies of the Witt algebra in its ideal part. Furthermore, the \Max\ algebra can be formally deformed into two new families of algebras when we consider deformations of other commutators. The new infinite dimensional algebras obtained have been denoted as $M(a,b;c,d)$ and $\bar{M}(\bar{\alpha},\bar{\beta};\bar{\nu})$.
In particular, the \Max\ algebra can be formally deformed to the (twisted) Schrödinger-Virasoro algebra for the specific values of parameters $a=c=d=0$ and $b=-\frac{1}{2}$, which can be seen as the asymptotic symmetry algebra of the spacetimes invariant under Schrödinger symmetry \cite{Alishahiha:2009nm, Compere:2009qm}. %\textcolor{red}{the asymptotic symmetry of Schr\"{o}dinger spacetimes with the specific values of parameters as $M(\frac{z-2}{2z},\frac{-1}{z};\frac{z-2}{z},\frac{z-2}{z})$ \cite{Compere:2009qm}. This algebra is called (twisted) Schr\"{o}dinger-Virasoro algebra when one sets $z=2$ which is studied as the asymptotic symmetry algebra of the spacetimes invariant under Schr\"{o}dinger symmetry \cite{Alishahiha:2009nm}}. \hnote{I changed some sentences. If you think it is not necessary you can erase it. }
%\pnote{I really like the modifications but, in my opinion, we should try to avoid changes that are not necessary and required by the referee. So I have erase it and put the previous version.}\hnote{I agree.}

We have then considered possible central terms for the obtained algebras through deformation procedure. We have first briefly review the well-known central extensions of the  $\mathfrak{bms}_{3}$ and the $\mathfrak{witt}$ algebra. We also explored the central extensions of $M(a,b;c,d)$ and $\bar{M}(\bar{\alpha},\bar{\beta};\bar{\nu})$ in some specific points of their parameters space. For a generic point in the parameter space $M(a,b;c,d)$ algebra  admits only one central term in its Witt subalgebra. For specific values of parameters it can admit more central terms which means that the deformation procedure can change the number of possible non trivial central terms. On the other hand the algebra $\bar{M}(\bar{\alpha},\bar{\beta};\bar{\nu})$ in general admits two non trivial central terms and a third central terms can appear for $\bar{\alpha}=\bar{\nu}=0$ in $\bar{M}(\bar{\alpha},\bar{\beta};\bar{\nu})$ as in the \Max\ algebra.

It is important to emphasize that two family algebras $M(a,b;c,d)$ and $\bar{M}(\bar{\alpha},\bar{\beta};\bar{\nu})$ have been obtained by deforming commutators being not at the level of the ideal part. Interestingly, similar results have been obtained by deforming the \bms\ and $\mathfrak{bms}_{4}$ algebras in \cite{Parsa:2018kys, Safari:2019zmc}. The examples considered in this paper, hence confirm the conjecture made in \cite{Parsa:2018kys,Safari:2019zmc}, that the Hochschild-Serre factorization (HSF) theorem\footnote{The Hochschild-Serre factorization (HSF) theorem states that we can only deform the ideal part of Lie algebra and other commutators remain untouched.} might be extended for infinite dimensional algebras as follows: the infinite dimensional Lie algebra\footnote{Here by infinite dimensional Lie algebras, we mean those algebras who are obtained as extensions of the Witt algebra.} with countable basis can be deformed in all of its commutators but only by terms with coefficients from the ideal part. The results obtained for the \Max\ algebra reinforce this conjecture.
%\pnote{In my opinion, I prefer to maintain the three first paragraphs since the referee has observations for the last 5 paragraph and it could be confusing for him to change the order of paragraph}\hnote{I agree. }

It is interesting to point out that the central extension of one of our deformations of the $\mathfrak{Max}_3$ algebra is a known asymptotic symmetry. Indeed three copies of the Virasoro algebra describes the asymptotic structure of a three-dimensional Chern-Simons gravity theory invariant under the so-called AdS-Lorentz algebra \cite{Concha:2018jjj}. In the stationary configuration, analogously to the Maxwell case, the additional gauge field appearing in the AdS-Lorentz case modifies the total energy and angular momentum. It would be interesting to explore how the total energy and angular momentum are influenced by the additional gauge field related to the other deformations and analyze the existence of a limit allowing to recover the conserved charges of the Maxwell one or those of General Relativity. Moreover, the study of a limit allowing to recover known gravity theories from a CS action based on the deformations considered here could be of interest. In particular, if a gravity theory based on an enlarge symmetry is appropriate for approach more realistic theories then these theories should at least satisfy the correspondence principle, namely they must be related to General Relativity.

On the other hand, as was discussed in \cite{Chernyavsky:2020fqs}, there is a particular choice of the parameters appearing in the Hietarinta-Maxwell Chern-Simons gravity theory which do not break the Hietarinta-Maxwell algebra but deforms it to three copies of the $\mathfrak{sl}(2,R)$ algebra which coincide with the finite subalgebra of the three copies of the $\mathfrak{witt}$ algebra. Then it would be interesting to explore if there is other choice of the parameters of the Hietarinta-Maxwell theory leading to the finite dimensional deformations presented here. Regarding the central extension of the deformed \Max\ algebra, one could study if there is a particular range of the parameters appearing in the central charges obtained here allowing to reproduce those of known theories.

%Let us note that our results can also be seen as all the possible deformations of the infinite dimensional enhancement of the simplest Hietarinta algebra \cite{Hietarinta:1975fu}. Such symmetry is obtained by interchanging the role of the generators of the ideal part of the Maxwell symmetry. All the deformations presented here then correspond to the deformations of the infinite dimensional Hietarinta algebra by interchanging the generator $\mathcal{M}_m$ with the generator $\mathcal{P}_m$. Further developments of this dual version of the Maxwell algebra have been recently presented in \cite{Bansal:2018qyz, Chernyavsky:2019hyp}.

Another aspect that deserves to be explored is the explicit derivation of the infinite-dimensional algebras introduced here by considering suitable boundary conditions. One could conjecture that the deformations of the $\mathfrak{Max}_3$ algebra should correspond to the respective asymptotic symmetries of three-dimensional Chern-Simons gravity theories based on deformations of the Maxwell algebra. %\textcolor{red}{As the specific case it is interestiong to explore $\mathfrak{bms_3}\oplus \mathfrak{witt}$ algebra as asymptotic symmetry algebra of the Chern-Simons gravity invariant under $\mathfrak{iso}(2,1)\oplus\mathfrak{so}(2,1)$ \cite{bms3-witt}.}
%Such conjecture has the following origin: As was shown in \cite{Parsa:2018kys} the \Max\ algebra appears as \textcolor{red}{an extension} \textcolor{blue} {and deformation} of the $\mathfrak{bms_3}$ algebra which result to be the corresponding asymptotic symmetry of a Chern-Simons gravity theory invariant under \textcolor{red}{an extension} \textcolor{blue} {and deformation} of the Poincaré algebra known as the Maxwell algebra \cite{Concha:2018zeb}. %Furthermore, one of our results supports this idea since we have shown that a deformation of the \Max\ algebra describes the asymptotic symmetry of a three-dimensional CS gravity theory invariant under a particular deformation of the Maxwell algebra denotes as AdS-Lorentz algebra. 
Naturally, one could obtain a large number of possible asymptotic symmetries for several CS gravity models. The physical implications and motivations of every deformation should properly studied first. As in the Maxwell case (or in the Hietarinta case), the new deformations could have interesting features which would be worth it to study. In particular, one could explore if the theory invariant under deformation of the Maxwell algebra may change the thermodynamics properties of Black hole solution like their entropy.%\pnote{I have erased the sentence about poincaré + lorentz because, thinking as the referee, we do not provide any motivation to explore bms3+witt algebra, we just say that it is interesting but without any reason.}

It is worthwhile to study possible generalizations of our results to other (super)symmetries. The study of the solutions and asymptotic structure of the Maxwell superalgebra and its deformations remains as an interesting open issue. %An extension and deformation of the $\mathfrak{Max}_3$ algebra has been introduced in \cite{Caroca:2017onr} corresponding to the infinite enhancement of a generalized Maxwell algebra, also called $\mathfrak{B}_5$ algebra. In particular, it would be interesting to study possible deformations of such infinite enhancement. One could conjecture that the deformations would reproduce $\mathfrak{witt}\oplus\mathfrak{witt}\oplus\mathfrak{bms}_3$ algebra, $\mathfrak{witt}\oplus\mathfrak{Max}_3$ algebra or some generalization of the family algebras $M(a,b;c,d)$ and $\bar{M}(\bar{\alpha},\bar{\beta};\bar{\nu})$. At the supersymmetric level, one could explore all possible deformations of the infinite dimensional enhancement of the $\mathcal{N}$-extended Maxwell superalgebra recently introduced in \cite{Caroca:2019dds}. 
Furthermore, one could analyze for which values of the parameters the family algebras $M(a,b;c,d)$ and $\bar{M}(\bar{\alpha},\bar{\beta};\bar{\nu})$ admit a well-defined supersymmetric extension. %One might obtain them through a deformation procedure from the supersymmetric extension of the \Max\ algebra presented recently in \cite{Caroca:2019dds}. The same study could be extended to the family algebra $W(a,b)$ which appears as a deformation of the \bms\ algebra \cite{Parsa:2018kys}. 
The next problem which would be interesting to explore is studying the group associated to the \Max\ algebra and asking how deformation procedure affects at the group level and its representations. Recently, the group associated to \Max\ algebra and its coadjoint orbits have been considered \cite{Salgado-Rebolledo:2019kft} so one might asked about the connection between coadjoint orbits of this group and the groups associated to the deformation of \Max\ obtained here. %such as three copies of Witt algebra or $\mathfrak{bms}_{3}\oplus\mathfrak{witt}$ or two family algebras $M(a,b;c,d)$ and $\bar{M}(\bar{\alpha},\bar{\beta};\bar{\nu})$.
%In other words one may explore how deformation relates the Hilbert spaces and unitary representations of two groups (algebras) which are connected by deformation procedure.
%\pnote{I have reduced considerably the content of the two last paragraphs and put all the content in a unique last paragraph. In this way we avoid, excesive references and statements}

%\hnote{We should also discuss about physical interpretations of deformation parameters. For instance, the parameter $a,b,c,d$ can be interpreted as parameter which take us from Maxwell algebra (space) to Schrodinger algebra (space). }
%%%%%%%%%%%%%%%%%%%%%%%%%%%%%%%%%%%%%%%%%%%%%%%%%%%%%%%%%%%%%%%%%%%%%%%%%%%%%%%%%%%%%%%%%%%%%%%%%%%%%%%%%%
\section*{Acknowledgement}
This work was supported by the CONICYT - PAI grant No. 77190078 (P.C.). H. R. S acknowledge the partial support of Iranian NSF under grant No. 950124.
H.R.S. wishes to thank to M. Henneaux and S. Detournay for their kind hospitality at the Physique Théorique et Mathématique department of the Université Libre de Bruxelles (ULB) which part of this work was done and acknowledges the support by F.R.S.-FNRS fellowship "Bourse de séjour scientifique IN". P.C. would like to thank to the Dirección de Investigación and Vice-rectoria de Investigación of the Universidad Católica de la Santísima Concepción, Chile, for their constant support. The authors would like to thank H. Afshar, S. Detournay and M. M. Sheikh-Jabbari for the correspondence and comments.

\bibliographystyle{fullsort.bst}
 
\providecommand{\href}[2]{#2}\begingroup\raggedright\endgroup

\end{document}